\documentclass[usenatbib,fleqn]{mn2e}

\usepackage{graphicx}
\usepackage{xspace}
\usepackage{latexsym}
\usepackage{amssymb}
\usepackage{amsmath}
\usepackage{journals}
\usepackage{url}

\newcommand{\eqn}[1]{Eq.~(\ref{#1})}

\newcommand{\fgr}[1]{Fig.~\ref{#1}}
\newcommand{\mrm}{\mathrm}

\newcommand{\X}{\textsf{X}\xspace}
\newcommand{\Z}{\textsf{Z}\xspace}
\newcommand{\BH}{BH\xspace}
\newcommand{\BHB}{BHB\xspace}
\newcommand{\m}{-}
\newcommand{\D}{\mathnormal{d}}
\newcommand{\DS}{\displaystyle}
\newcommand{\FW}{1.0\columnwidth}

\def\beq{\begin{equation}}
\def\eeq{\end{equation}}
\def\beqa{\begin{eqnarray}}
\def\eeqa{\end{eqnarray}}

\defcitealias{zier06}{Paper~I}

\title[Merging of a massive black hole binary II]{Merging of a massive black
  hole binary II}

\author[C. Zier]{C. Zier$^{1}$\thanks{E-mail: chzier@mpifr-bonn.mpg.de}\\
$^{1}$Raman Research Institute, Bangalore 560080, India}
\begin{document}

\date{Accepted Received}

\pagerange{\pageref{firstpage}--\pageref{lastpage}} \pubyear{2006}

\maketitle

\label{firstpage}

\begin{abstract}
In this paper, the second in a series of two, we justify two important
assumptions on which the result is based that in course of a galaxy
merger the slingshot ejection of bound stars is sufficiently efficient
to allow a supermassive black hole binary to merge. A steep cusp with
a power law index of $2.5\text{--}3$ is required which is as massive
as the binary and surrounds the \BH{}s when the binary becomes
hard. This cusp is probably formed when both clusters, surrounding
each black hole, merge and combine with the matter funneled into the
center. We find this profile to be in agreement with observed
post-merger distributions after the cusp has been destroyed. The time
dependency we derive for the merger predicts that stalled black holes,
if they exist at all, will preferably be found in less than $\sim
0.2\,\mrm{pc}$ distance. To test this prediction we compute the
current semimajor axis of 12 candidates of ongoing mergers. We find
all binaries unambiguously to be already in the last phase when they
decay due to the emission of gravitational waves. Therefore, in
striking contradiction with predictions of a depleted loss-cone, the
abscence of even a single source in the slingshot phase strongly
supports our previous and current results: Binaries merge due to
slingshot ejection of stars which have been funneled into the central
regions in course of a galaxy collision.
\end{abstract}


\section{Introduction}
\label{s_intro}
The formation of a supermassive black hole binary (BHB) is the natural
consequence of two widely accepted assumptions: Galaxies harbour a
supermassive black hole (BH) in their center and galaxies merge with
each other. Such \BHB{}s, merged and not yet merged, are important
because they are used to explain a wide variety of features observed
in galaxies. For a detailed review on observational evidence of
\BHB{}s see \citet{komossa03b} or an updated version,
\citet{komossa06}. The evolution of the merging \BH{}s can be
subdivided into three successive phases \citep{begelman80}: In the
beginning both cores spiral inwards to their common center due to
dynamical friction. Once the \BH{}s bind to each other on the parsec
scale and form a hard binary they keep on merging due to slingshot
ejection of stars. Finally, in the third phase, the binary continues
to decay owing to the emission of gravitational waves. While the first
and third phase are well investigated it is still a matter of debate
whether the slingshot ejection of stars in the second phase is
efficient enough to enable the binary to enter the final phase or
whether the merging process comes to a halt due to loss-cone
depletion. Even though numerical scattering experiments showed that
the \BH{}s merge on scales of $10^{8\text{--}9}\,\mrm{yr}$
\citep{quinlan96,makino97,quinlan97,milos01,zier01} it is argued in
all publications but the last that the loss-cone becomes depleted long
before the binary enters the third phase and the binary probably gets
stalled. This reasoning is based on the assumption that the binary is
embedded in a flat spherically symmetric core which is derived from
the central density profiles of elliptical galaxies
\citep{berczik05}. According to hierarchical models for galaxy
formation this type of galaxy has experienced a major merger
previously and therefore its mass has been redistributed from the
central parts to the outer regions, resulting in a flat profile after
the merger.

While there is no conclusive observational evidence for stalled
binaries various sources suggest a generally successful merger of the
\BH{}s. \citet{haehnelt02} argued that if the merging time would
exceed a Hubble time the binary should become ejected in about 40\% of
bright ellipticals when merging with a third galaxy. They pointed out
that this would be in contradiction with the \BH{}s which have been
observed indirectly in all nearby elliptical galaxies and with the
small scatter of the $M_\mrm{BH}\, \text{--}\, \sigma_\ast$ relation
\citep[e.g.][]{gebhardt00,tremaine02}. A certain class of sources, the
so-called \X-shaped radio galaxies (XRGs), can be well explained in
terms of a completed merger of a \BHB, an interpretation first used by
\citet{rottmann01}. When the \BH{}s finally coalesce the spin axis is
rapidly realigned into the direction of the orbital angular momentum
so that the old and new lobes appear as an \X on the sky
\citep{zier01,dennett-thorpe02,zier02,gergely07}. During the third
phase the rapidly precessing jet produces effectively a powerful wind,
which entrains the environmental gas and is identified by
\citet{gergely07} with a superdisc, dicussed by \citet{gopal07}. A
merger is also held responsible for \Z-shaped radio galaxies, where
the secondary galaxy bends the jet of the primary into a \Z-shape
before the \BH{}s coalesce \citep{gopal-krishna03,zier05}, for
double-double radio galaxies \citep{schoenmakers00,liu03} and possibly
for compact symmetric objects \citep{zier02}. Recently the sample of
known XRGs has been increased considerably by \citet{cheung07}. This
can be used for systematic studies and hence might support the merging
scenario as formation mechanism of XRGs.  Other sources suggest that
the \BH{}s have not yet merged and are still orbiting around each
other. Helical jet patterns could be explained in this way
\citep{begelman80} as well as semi-periodic changes in lightcurves
\citep[e.g.][]{sillanpaa88,katz97}. However, this does not necessarily
mean that the \BH{}s are stalled, the binary might still decay.

In a recent letter \citep[from now on \citetalias{zier06}]{zier06} we
showed that the slingshot ejection of stars in the second phase is
efficient enough to allow the \BH{}s to shrink to the third phase and
coalesce within less than a Hubble time. Unlike in previous numerical
simulations where the focus was on stars scattered off the binary, we
focused on the stars bound in the potential of the \BH{}s. The
results showed that if the binary by the time it becomes hard is
surrounded by a flat cusp with a power law index $\gamma\lesssim 2$,
as it appears \emph{after} the merger and has been used in previous
simulations, it will stall in this phase unless the cusp is very
massive. However, we predict that the cusp is as massive as the binary
and sufficiently steep ($\gamma\gtrsim 2.5$) \emph{during} the merger
when the binary becomes hard. The ejection of this mass out of the
potential of the \BH{}s extracts enough energy so that the binary can
enter the third phase and the \BH{}s coalesce. We argued that such a
profile is formed during the merger. Parameters like the initial mass
and velocity distributions in the isolated galaxies as well as the
magnitude and orientation of both galactic spins and the orbital
angular momentum relative to each other have a strong influence on the
merger and the morphology of the remnant \citep{toomre72}. While the
galaxies are merging energy is dissipated and angular momentum
redistributed with some fractions compensating each other. Large
amounts of mass move on highly eccentric orbits \citep{rauch96} in a
potential that is stronlgy non-spherically symmetric. Low angular
momentum matter accumulates in the center. Together with both cores
which surround each \BH and whose density increases considerably
during the merger \citep{barnes96} this matter forms a massive and
steep cusp by the time the binary becomes hard. This cusp is only
transient because it will be destroyed by the merging binary and
therefore is not likely to be observed.  However, this should be the
appropriate profile in order to simulate the second phase.

In the present article, after repeating in
Section~\ref{s_preliminaries} the results from \citetalias{zier06}
which we will need in this paper, we will justify in more detail our
assumptions for the kick-parameter $k$ (Section~\ref{s_k}) and the
neglegt of the cluster potential
(Section~\ref{s_potential}). Afterwards we show that the fraction of
mass which is required to become ejected is in agreement with the
total mass of the galaxy. We also show that the profile which at the
beginning of the second phase is required to allow the \BH{}s to
coalesce is in agreement with the observed post-merger profiles after
the binary has destroyed the cusp (Section~\ref{s_density}). In
Section~\ref{s_shrinking-rate} we briefly consider the evolution of a
merger in time before we compare the effects of multiple mergers on
the ejected mass with numerical simulations in
Section~\ref{s_multiple}. Afterwards we examine observational evidence
for ongoing mergers which might have become stalled
(Section~\ref{s_ongoing}) and finally summarize our results in
Section~\ref{s_summary}.


\section{Preliminaries}
\label{s_preliminaries}
First we repeat the basic assumptions and some results from
\citetalias{zier06} which we will use in the present article. It is
assumed that the \BH{}s, moving on Keplerian orbits, have formed a
hard binary and that the origin coincides with the center of mass. We
define the mass ratio $q\equiv m_2/m_1 \leq 1$. The total and reduced
mass are $M_{12} = m_1 + m_2$ and $\mu = m_1 m_2/M_{12}$,
respectively. Hence, the energy of the binary is
\begin{equation}
E_\mrm{bin} = -\frac{GM_{12}\mu}{2a}
\label{eq_e-bin}
\end{equation}
and the relative velocity between the BH{}s is
\begin{equation}
v_\mu = \sqrt{\frac{GM_{12}}{a}},
\label{eq_v-mu}
\end{equation}
where $a$ is the semimajor axis of the binary. If the cluster mass $M_\mrm{c}$
is distributed according to the power law $\rho = \rho_0 (r/r_0)^{-\gamma}$
between the radii $r_\mrm{i}$ and $r_\mrm{c}$ with 
\begin{equation}
\rho_0 = \frac{M_\mrm{c}}{4\pi r_0^3} \begin{cases}
  \frac{\DS 3-\gamma}{\DS (r_\mrm{c}/r_0)^{3-\gamma}
  -(r_\mrm{i}/r_0)^{3-\gamma}}, & \gamma \neq 3\\[2.5ex]
  \frac{\DS 1}{\DS \ln(r_\mrm{c}/r_\mrm{i})}, & \gamma = 3.
\end{cases}
\label{eq_rho0}
\end{equation}
we obtain for the mass within $r$
\begin{equation}
M(r) = M_\mrm{c} \begin{cases} \frac{\DS r^{3-\gamma} -
  r_\mrm{i}^{3-\gamma}}{\DS r_\mrm{c}^{3-\gamma} - r_\mrm{i}^{3-\gamma}}, &
  \gamma \neq 3\\[2.5ex] \frac{\DS \ln(r/r_\mrm{i})}{\DS
  \ln(r_\mrm{c}/r_\mrm{i})}, & \gamma = 3.
\end{cases}
\label{eq_m}
\end{equation}
Note that in the following we will refer to the logarithmic slopes of
density profiles, $\D\log\rho/\D\log r$, simply as slopes. In
\citetalias{zier06}, we showed that a mass of about $2 M_{12}$ is bound
to the binary, of which a large fraction is expected to be in the
loss-cone. For stars belonging to this population, the initial energy
in the potential of the binary is
\begin{equation}
E_{\ast,\mrm{i}} = -(1-\epsilon)\frac{G M_{12}m_\ast}{2 r_\m},
\label{eq_e-i}
\end{equation}
where $\epsilon <1$ is the eccentricity of the star's orbit and $r_\m$ the
pericenter. We approximated the potential of the binary to first order with a
point potential of mass $M_{12}$ located at the center of the cluster.  This
introduces only minor deviations with a maximum of a factor of less than $2$
for $q=1$. In Section~\ref{s_potential}, we will show that it is justified to
neglect the potential of the cluster itself when computing the binding energy
of the stars in \eqn{eq_e-i}. After its ejection the formerly bound star will
have a positive energy which we can scale with the factor $\kappa$ to its
initial energy for circular orbits ($\epsilon = 0$):
\begin{equation}
E_{\ast,\mrm{f}} = \kappa\frac{G M_{12}m_\ast}{2 r_\m}.
\label{eq_e-f}
\end{equation}
According to \citet{quinlan96} the dominant contribution to the hardening of
the binary comes from stars whose pericenter is about the semimajor axis of
the binary, independent of the density profile. Hence replacing $r_\m$ with
$a$ we obtain for the energy change $E_{\ast,\mrm{f}} - E_{\ast,\mrm{i}}$ of
the star:
\begin{equation}
\Delta E_\ast = (1-\epsilon +\kappa) \frac{G M_{12}m_\ast}{2a} \equiv
k\,\frac{G M_{12}m_\ast}{2a} = k\frac{m_\ast v_\mu^2}{2}.
\label{eq_de-star}
\end{equation}
Note that for pericenters smaller than $a$ the initial energy of the
star would be smaller and therefore the energy change larger,
resulting in an increased kick-parameter $k$.

In the limit $m_\ast \ll m_2$ we can replace $m_\ast$ with $\D m$ and write
\eqn{eq_de-star} in its infinitesimal form. Equating it with the change of the
binary's energy in \eqn{eq_e-bin} due to the ejection of the mass $\D m$
yields the differential equation
\begin{equation}
\frac{\D a}{a} = -k \frac{\D m}{\mu},
\label{eq_da}
\end{equation}
which relates the shrinking of the binary to the amount of the ejected mass.
Integrating \eqn{eq_da} from $a_\mrm{g}$ to $a_\mrm{h}$ we showed in
\citetalias{zier06} that the binary has to eject the mass
\begin{equation}
m_\mrm{ej} = \frac{\mu}{k} \ln \frac{a_\mrm{h}}{a_\mrm{g}}.
\label{eq_mej}
\end{equation}
The semimajor axis where the binary becomes hard, $a_\mrm{h}$, and
where emission of gravitational waves starts to dominate the decay,
$a_\mrm{g}$, mark the transitions from phase 1 to 2 and phase 2 to 3,
respectively. While $a_\mrm{g}$ is well defined (see \eqn{eq_ag}),
there is no unique prescription for the semimajor axis where the
binary becomes hard. However, the ratio of these distances $\eta\equiv
a_\mrm{h}/a_\mrm{g}$ is agreed to range from about $20$ to $100$, see
the discussion in \citetalias{zier06} and references therein. In the
following we assume that the binary is hard and do not worry about the
exact value of $a_\mrm{h}$. We showed that the ejection of about
$m_\mrm{ej}$ is sufficient for the binary to shrink from $a_\mrm{h}$
to $a_\mrm{g}$, provided this mass is distributed according to a steep
power law with an index $\gamma \gtrsim 2.5$. Therefore, we concluded,
the coalescence of the \BH{}s is very likely in the course of a galaxy
merger where a large amount of mass with low angular momentum is
accumulated in the central region.


\section{The kick-parameter \lowercase{$k$}}
\label{s_k}
One of our basic assumptions in \citetalias{zier06} is a
kick-parameter $k=1$. In the literature we can find various
prescriptions for $k$ which we defined in \eqn{eq_de-star}. If we
express the final energy of the ejected star in terms of its velocity
at infinity, $E_{\ast,\mrm{f}} = m_\ast v_\infty^2/2$ we can write
$\kappa = (v_\infty/v_\mu)^2$, where we made use of
Eqs.~(\ref{eq_e-f}) and (\ref{eq_v-mu}) and replaced the pericenter of
the orbit of the star with the semimajor axis of the binary. This form
can be used in the relation between the scaling parameters obtained
from \eqn{eq_de-star}, $k = 1-\epsilon + \kappa$. From scattering
experiments \citet{quinlan96} finds that most of the stars are ejected
with a final velocity $v_\infty \approx (3/2)v_\mu \sqrt{m_2/M_{12}} =
(3/2) v_\mu\sqrt{q/(1+q)}$ and hence we obtain
\begin{equation}
k = 1-\epsilon + \frac{9}{4}\,\frac{q}{1+q}.
\label{eq_k-quinlan}
\end{equation}
\citeauthor{quinlan96} argued that the energy gained by a star can
basically be attributed to the interaction with the smaller \BH in the
limit $m_1 \gg m_2$, because the larger \BH acts as a fixed
potential. He then derives an expression for the energy change which
is proportional to $m_2/M_{12} = q/(1+q)$. However, because $m_1$ and
$m_2$ are bound to each other also the acting forces correspond to
each other so that the larger mass $m_1$ compensates for the smaller
semimajor axis of its orbit. Therefore it does not seem to be
justified to approximate it as a fixed point potential compared to the
potential generated by $m_2$. The potential of the \BH{}s is $\Phi_i =
Gm_i/r_i$. As they move along their orbits a test mass $m_\ast$, which
is fixed in space, experiences a change in the potential which is
proportional to the displacement of the \BH{}s:
\begin{equation}
\D \Phi_i = \frac{Gm_i}{r_i^2} dr_i .
\end{equation}
We assume that the displacement $dr_i$ of the mass $m_i$ corresponds to the
semimajor axis of its orbit $a_i$. Expressing this in terms of the semimajor
axis of the binary $a$, i.e. $a_1= a q/(1+q)$ and $a_2= a/(1+q)$, we can write
\begin{equation}
\D \Phi_1 = \frac{Gm_1}{r_1^2} a \frac{q}{1+q},\qquad \D \Phi_2 =
\frac{Gm_2}{r_2^2} a \frac{1}{1+q}.
\end{equation}
The binary shrinks mostly due to the interaction with stars whose
closest approach corresponds to the semimajor axis
\citep{quinlan96}. Hence we can replace $r_1$ and $r_2$ with $a$,
resulting in equal changes of both potentials, $\Delta\Phi =
(Gm_1/a)\, q/(1+q) = v_\mu^2 \,\mu/M_{12}$. If the star was moving on
a parabolic orbit and this energy is tranfered from both \BH{}s to the
star its final velocity is $v_\infty = v_\mu\sqrt{2\mu/M_{12}}$ so
that
\begin{equation}
\kappa = 2 \frac{\mu}{M_{12}} = 2 \frac{q}{(1+q)^2}.
\label{eq_kappa-pot}
\end{equation}
This result has been cited before to have been obtained by
\citet{saslaw74} in numerical experiments. Apart from a factor
$1/(1+q)$, which has a minimum of $1/2$ for $q=1$, this result is very
similar to that obtained by \citeauthor{quinlan96} above. Similar
values have also been found before. In simulations of close encounters
of stars with a hard equal-mass binary of zero eccentricity,
\citet{hills80} obtained for the mean velocity of stars at infinity
$v_\infty \approx 0.84\, v_\mu$, and hence $\kappa \approx
0.71$. Later \citet{roos81} performed numerical computations with a
varying mass ratio and approximated the kick-parameter with $k=2
\mu/M_{12}$, the same result we have found above for $\kappa$. For
parabolic orbits, as we have assumed above, these parameters are equal
and our crude estimate is in good agreement with his result.

More recently \citet{zier00} simulated a stellar cluster bound in the
potential of a \BHB which is moving on fixed circular orbits. He
carried out several runs for different mass ratios ($q=0.01$, $0.1$
and $1$) and various initial density profiles of the stars (Gaussian
or power laws with index $\gamma = 2$ or $4$). For every run we binned
the initial eccentricities of the orbits and computed the
kick-parameter for each bin. In \fgr{f_k-eps}, we show the thus
obtained $k$ in dependency on the eccentricity and find a distinct
linear correlation. We can not detect a clear dependency on the kind
of the initial profile in the plot. However, the data show a weak
positive dependency of $k$ on the mass ratio $q$. This is in agreement
with the previous results given above in Eqs.~(\ref{eq_k-quinlan}) and
(\ref{eq_kappa-pot}) if we write $k=1-\epsilon + \kappa$, although the
dependency on $q$ exhibited in the data is weaker. However, we find
that $k\approx 1$ is a good approximation for $\epsilon \lesssim
0.4$. Note that the simulations by \citet{zier00} did not take into
account the potential of the stellar cluster and hence the real values
for $k$ should actually be larger than those displayed in
\fgr{f_k-eps}, see next Section. Because \eqn{eq_k-quinlan} is less
steep than the data in \fgr{f_k-eps} it suggests this equation
generally yields slightly larger values for the kick-parameter if the
mass ratio is not too small.

Comparing our definition of the kick-parameter with that of
\citet{yu02} for $K$ we find $k=2K\, q/(1+q)^2$. Making use of the
results of \citet{quinlan96}, \citeauthor{yu02} obtains $K\approx
1.6$. This translates to a maximum value of $k=0.8$ for $q=1$, roughly
in agreement with the previous results. Note that in
\citetalias{zier06} we did not include the factor $\mu/M_{12} =
q/(1+q)^2$ and so derived a too large value of $k=3.2$. Because we
just quoted this result and used $k=1$ throughout the paper none of
the results and conclusions obtained there are affected.
\begin{figure}
\begin{center}
\includegraphics[width=\FW]{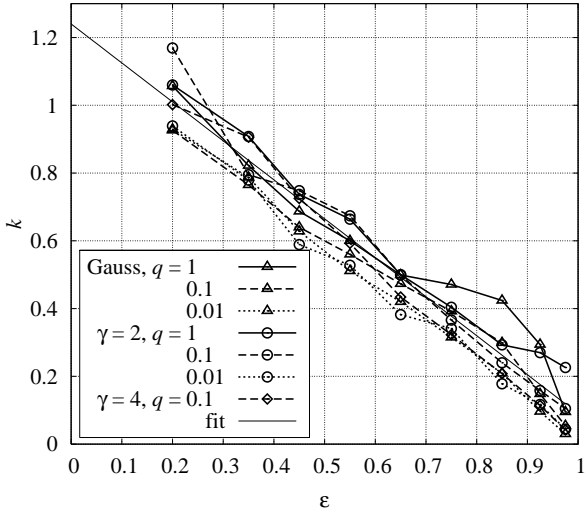}
\caption[]{Data show that the kick-parameter is roughly a linear function of
  the eccentricity of the stellar orbit. While $k$ tends to increase with $q$
  it does not seem to depend on the choice of the initial profile. The fit is
  drawn by eye using all data points.}
\label{f_k-eps} 
\end{center}
\end{figure}

In \eqn{eq_de-star} we defined $k$ after having replaced the pericenter of the
star $r_\m$ with the semimajor axis $a$ of the binary. Because only stars with
$r_\m\lesssim a$ can interact with the binary and become ejected and a
pericenter less then $a$ would increase the kick-parameter (keeping the
eccentricity constant) the values we derived should actually be a lower
limit. In conclusion we can say that the above results clearly show that $k$
is of order of $1$ unless the stars are moving on very eccentric orbits
(keeping $r_\m = a$ constant, i.e. stars are only weakly bound, what is very
unlikely due to dissipation of energy during the merger) and the mass ratio is
very small. Therefore our choice of $k=1$ in \citetalias{zier06} was well
justified and we continue to use this value in the present paper.


\subsection{The influence of the cluster potential on $k$}
\label{s_k-potential}
After a star interacted with the binary it will be ejected from the
potential of the \BH{}s. This might not happen after the first
interaction, but ultimately it will be ejected unless before the next
encounter with the binary the pericenter is shifted due to star-star
interactions to distances too large as to interact with the binary. Or
the binary has shrunk in the meantime to a semimajor axis much smaller
than the pericenter of the star, again resulting in no more
interactions. However, this will happen most likely only in the
beginning of the merger when the evolution is fastest (see
Section~\ref{s_shrinking-rate}) to stars whose energy is close to zero
after the last interaction so that they have been almost ejected
anyway.

Interacting with the binary some stars will be accelerated to a speed
which exceeds the escape velocity of the binary, but is less than the
escape velocity of the combined potentials of the \BH{}s and the
cluster. These stars stay bound to the center. After multiple
interactions with the binary the fraction of stars whose pericenters
have become larger relative to the semimajor axis of the binary for
the reasons given above will remain bound in the cluster without
interacting with the binary anymore. On longer time scales they might
diffuse back into the loss-cone. The other fraction eventually becomes
ejected from the total potential of the \BH{}s and cluster after
multiple encounters. Hence this delayed ejected fraction, emerging
because of the inclusion of the cluster potential, increases the
kick-parameter on average. At radii $r\geq r_\mrm{c}$ the star's
specific energy is
\begin{equation}
\begin{split}
{\cal E}_\ast & = & \frac{v^2}{2} - \frac{G(M_{12}+M_\mrm{c})}{r}\\ & = &
\frac{v^2}{2} - \frac{GM_{12}}{r}(1+f),
\end{split}
\label{eq_e-star-c}
\end{equation}
where $f\equiv M_\mrm{c}/M_{12}$, with $f>1$. The escape velocity in
the combined potential at $r>r_\mrm{c}$ is $v_\mrm{esc}^2 = 2(1+f)\,
GM_{12}/r$ and for the velocity of the star in the potential of the
binary only we can write $v^2(r) = 2({\cal E}_\ast + GM_{12}/r)$. If
we require that this velocity is at least as large as the escape
velocity at $r=r_\mrm{c}$ we obtain for the specific energy the
relation ${\cal E}_\ast \geq f\,GM_{12}/r_\mrm{c}$. Using this again
in the expression for the star's velocity in the limit of an infinite
radius yields $v_\infty^2 = 2{\cal E}_\ast \geq
2f\,v_\mu^2\,a/r_\mrm{c}$. Therefore the condition that the stars
become ejected from the combined potential of the binary and cluster
can be written as
\begin{equation}
\kappa \geq 2f\frac{a}{r_\mrm{c}}.
\label{eq_kappa}
\end{equation}
$\kappa$ is determined using only the binary potential. For a star
which eventually escapes from the binary and the cluster this
parameter is increasing with the normalized cluster mass $f$.  In
comparision with a neglected cluster potential ($f=0$)
\eqn{eq_kappa} shows that including this potential increases
$\kappa$. This results in a larger kick-parameter of the delayed
ejected fraction and therefore also of the mean value of $k$ of all
ejected stars. On the other hand we can use \eqn{eq_kappa} to derive
a maximum cluster mass for which stars will be ejected,
\begin{equation}
f \leq \frac{r_\mrm{c}}{a}\,\frac{\kappa}{2} = \frac{r_\mrm{c}}{a}\,
\frac{k+\epsilon -1}{2}.
\label{eq_l}
\end{equation}
With our results from the previous Section and assuming that
$r_\mrm{c}\gg a$ this relation still allows the cluster mass to exceed
the binary's mass by a factor of a few, as required by a successful
merger. Therefore most of the stars get a kick large enough to escape
from the center, even if we neglect the cluster
potential. Consequently the fraction of delayed ejected stars is
small, only slightly increasing the mean of $k$. This is in agreement
with \citet{yu02} who finds that especially for $a\ll a_\mrm{h}$ stars
generally escape from the binary and cluster.  We can summarize that
including the cluster potential tends to increase on average the value
of the parameter $k$ so that the values obtained above are rather
lower limits.


\section{The cluster potential}
\label{s_potential}
We also have to check whether it is justified to neglect the cluster
potential when computing the potential energy of stars bound by the
binary, our second basic assumption in \citetalias{zier06}. If
initially stars in the cluster are not bound by the \BH{}s, they are
at least bound by the cluster itself. The energy of a star before and
after the interaction with the \BH{}s is
\begin{align}
\label{eq_ini-egy}
E_{\ast,\mrm{i}} & = \frac{m}{2}v_\mrm{i}^2 - m (\Phi_\mrm{bin}
+\Phi_\mrm{c})\\
E_{\ast,\mrm{f}} & = \frac{m}{2}v_\mrm{f}^2 - m (\Phi_\mrm{bin} +\Phi_\mrm{c}),
\label{eq_fin-egy}
\end{align}
with $\Phi_\mrm{c}$ and $\Phi_\mrm{bin}$ being the potential of the
cluster and the binary, respectively. If we compare the initial and
final energy of the star at the same radius and assume that both
potentials did not change during the time of the interaction between
the binary and the star we can write for the change of the energy
\[
\Delta E_\ast = E_{\ast,\mrm{f}} - E_{\ast,\mrm{i}} = \frac{m}{2}(v_\mrm{f}^2
- v_\mrm{i}^2).
\]
Taking the circular velocity in a point potential as the typical velocity of a
star in the cluster we showed in \citetalias{zier06} that a mass of about
$2M_{12}$ is gravitationally bound to the binary. Depending on the power law
index of the mass distribution the radius of a sphere which contains twice the
mass of the binary is (\eqn{eq_m})
\begin{equation}
r_\mrm{b} = r_\mrm{i} \begin{cases} \left(1+2\frac{\DS M_{12}}{\DS
  M_\mrm{c}}\left[\left(\frac{\DS r_\mrm{c}}{\DS r_\mrm{i}} \right)^{3-\gamma}
  -1\right]\right)^{1/(3-\gamma)}, & \gamma \neq 3 \\[2.5ex] \left(\frac{\DS
  r_\mrm{c}}{\DS r_\mrm{i}}\right)^{2M_{12}/M_\mrm{c}} , & \gamma = 3.
\end{cases}
\label{eq_rb}
\end{equation}
For a valid solution of course the relation $M_\mrm{c} > 2 M_{12}$
must be satisfied. In this range the radius $r_\mrm{b}$ is increasing
with decreasing $\gamma$, i.e. larger for flatter profiles. For
$\gamma=2$ and $r_\mrm{i}\ll r_\mrm{c}$ we have $r_\mrm{b} =
2\,r_\mrm{c}\,M_{12}/M_\mrm{c}$, independent of $r_\mrm{i}$. If the
cluster is four times as massive as the binary and we assume
$r_\mrm{c}\approx 100\,\mrm{pc}$ we obtain for $r_\mrm{b}$ about
$50\,\mrm{pc}$, i.e. a radius much larger than the semimajor axis
$a_\mrm{h}$ where the binary becomes hard. For $\gamma = 3$ and
$M_\mrm{c}=4M_{12}$ we find $r_\mrm{b} = \sqrt{r_\mrm{i} r_\mrm{c}}$,
i.e. the geometrical mean. With $r_\mrm{i}=0.01$ and
$r_\mrm{c}=100\,\mrm{pc}$ this is $1\,\mrm{pc}$. Thus a mass of
$2M_{12}$ is contained in the central cusp and bound to the binary
when it becomes hard.

In \citetalias{zier06} we assumed the star to be bound to the binary,
i.e. $E_{\ast,\mrm{i}}$ is negative even if $\Phi_\mrm{c}$ is
neglected in \eqn{eq_ini-egy}. Then $v_\mrm{i}$ is less than the
escape velocity from the binary what is roughly true for stars within
the sphere of radius $\sim r_\mrm{b}$. It is this bound population on
which we focused in our previous paper and which we also consider in
the present work. When we calculated the energy of the stars and
derived the mass which is required to be ejected in order to extract
sufficient energy so that the BH{}s merge, we neglected the potential
of the cluster. This we will justify here by comparing the energy of a
star in the potential of the binary with its energy in the potential
of a stellar cluster whose mass is distributed according to a power
law with index $2$ or $3$.

In case of $\rho = \rho_0 (r/r_0)^{-2}$ we can write Poisson's equation as
\begin{equation}
\nabla^2\Phi_\mrm{c} = 4\pi G\rho = s/r^2,
\label{eq_poisson}
\end{equation}
where we have introduced the constant $s\equiv 4\pi G\rho_0^2$. The
stars in the cluster are distributed between $r_\mrm{i}$ and
$r_\mrm{c}$, the inner and the cluster radius, respectively. With
$M_\mrm{c}$ being the total mass of the cluster we can write $s =
GM_\mrm{c}/(r_\mrm{c} - r_\mrm{i})$. Integrating \eqn{eq_poisson}
twice we obtain for the potential in the range $r_\mrm{i} < r <
r_\mrm{c}$
\begin{equation}
\Phi_\mrm{c}(r) = \Phi_\mrm{c}(r_\mrm{i}) + s \left( \ln\frac{r}{r_\mrm{i}} +
\frac{r_\mrm{i}}{r} -1\right) - r_\mrm{i}^2 \left.\frac{\partial \Phi}{\partial
  r}\right|_{r_\mrm{i}} \left(\frac{1}{r} - \frac{1}{r_\mrm{i}}\right).
\end{equation}
Because the mass is spherically symmetric distributed and there is no
mass within $r_\mrm{i}$, no force is acting on a particle in this
range.  Therefore $F = -\partial \Phi/\partial r$ has to vanish at
$r=r_\mrm{i}$. On the other hand the force acting on a particle
outside the cluster is the same as that of a pointmass $M_\mrm{c}$
located at the origin, $-GM_\mrm{c}/r^2$ (Newton's second
theorem). Evaluating this condition at $r=r_\mrm{c}$ we can write the
potential in the form
\begin{equation}
\Phi_\mrm{c}(r) = - GM_\mrm{c} \begin{cases} \frac{\DS 1}{\DS r_\mrm{c} -
  r_\mrm{i}} \ln\frac{\DS r_\mrm{c}}{\DS r_\mrm{i}}, & 0\leq r\leq
  r_\mrm{i}\\[2.5ex] \frac{\DS 1}{\DS r_\mrm{c} - r_\mrm{i}} \left(1-\frac{\DS
  r_\mrm{i}}{\DS r} +\ln\frac{\DS r_\mrm{c}}{\DS r}\right), & r_\mrm{i}\leq r
  \leq r_\mrm{c}\\[2.5ex] \frac{\DS 1}{\DS r}, & r_\mrm{c}\leq r.
\end{cases}
\label{eq_pot-2}
\end{equation}
The potential energy of the cluster in its own potential is \citep[page
34]{binney94}
\begin{equation}
E_\mrm{c} = 2\pi \int r^2 \rho(r) \Phi(r) dr.
\label{eq_self-e}
\end{equation}
We are only interested in the stars which become ejected after interacting
with the binary, i.e. stars in the range between $a_\mrm{g}$ and
$a_\mrm{h}$. It is the energy of this fraction in the potentials of the
cluster and the binary that we want to compare. Therefore we have to integrate
\eqn{eq_self-e} in the limits from $a_\mrm{g}$ to $a_\mrm{h}$ and obtain
\begin{equation}
\begin{split}
E_\mrm{c} = & -\frac{GM_\mrm{c}^2}{2r_\mrm{c}} \frac{1}{(1-1/\zeta)^2}\\ &
\times \left[\frac{2}{\lambda}\,\frac{\eta -1}{\eta} -\frac{\ln\eta}{\zeta}
+\frac{\ln\lambda}{\lambda} -\frac{\ln(\lambda\eta)}{\lambda\eta}\right].
\label{eq_self-2}
\end{split}
\end{equation}
In this expression we used the following definitions
\begin{alignat}{3}
\label{eq_defs}
\eta & \equiv \frac{a_\mrm{h}}{a_\mrm{g}}, & \qquad\quad \zeta & \equiv
\frac{r_\mrm{c}}{r_\mrm{i}}, & \qquad\quad \lambda & \equiv
\frac{r_\mrm{c}}{a_\mrm{h}},\\
\intertext{which can be combined to}
\label{eq_ratios}
\frac{a_\mrm{h}}{r_\mrm{i}} & = \frac{\zeta}{\lambda}, & \qquad\quad
\frac{a_\mrm{g}}{r_\mrm{i}} & = \frac{\zeta}{\lambda\eta}, & \qquad\quad
\frac{a_\mrm{g}}{r_\mrm{c}} & = \frac{1}{\lambda\eta}.
\end{alignat}
The potential energy of this fraction of the cluster ($a_\mrm{g}\leq r \leq
a_\mrm{h}$) in the potential of the binary, which we approximated by that of a
pointmass $M_{12}$ at the origin, is
\begin{equation}
E_\mrm{bin} = -\frac{GM_{12}M_\mrm{c}}{r_\mrm{c} -
  r_\mrm{i}}\ln\frac{a_\mrm{h}}{a_\mrm{g}} =
  -\frac{GM_{12}M_\mrm{c}}{r_\mrm{c}} \frac{\ln{\eta}}{1-1/\zeta}.
\label{eq_e-p}
\end{equation}
This energy is obtained by multiplying the binding energy in Eq.~(11) of
\citetalias{zier06} with the factor $2$. We assume that the mass of the
cluster within the distance $a_\mrm{h}$ corresponds to $m_\mrm{ej}$, i.e.
\begin{equation}
M(a_\mrm{h}) = M_\mrm{c}\frac{a_\mrm{h}-r_\mrm{i}}{r_\mrm{c}-r_\mrm{i}} =
\frac{\mu}{k}\ln\frac{a_\mrm{h}}{a_\mrm{g}}.
\end{equation}
Solving for the cluster mass and using the definitions in \eqn{eq_defs} we
obtain
\begin{equation}
M_\mrm{c} = \frac{\mu}{k}\lambda\frac{\zeta - 1}{\zeta - \lambda} \ln\eta.
\end{equation}
As we argued in \citetalias{zier06} solutions with $r_\mrm{i} < a_\mrm{g}$ are
physically unreasonable because the mass in this range is not included in the
mass which is interacting with and ejected by the binary.  Therefore we assume
$r_\mrm{i} = a_\mrm{g}$ in the following, implying $\zeta = \lambda\eta$. The
ratio of the potential energies of the mass in the cusp range ($a_\mrm{g}\leq
r\leq a_\mrm{h}$) then results in
\begin{equation}
\frac{E_\mrm{c}}{E_\mrm{bin}} = \frac{1}{k}\, \frac{q}{(1+q)^2} \left(1 +
\frac{1}{2}\ln\lambda - \frac{\ln\eta}{\eta-1}\right).
\label{eq_e-ratio}
\end{equation}
This ratio is plotted in \fgr{f_e-ratio} as a function of $\lambda$
with the mass ratio $q$ as parameter (bold lines) and $k=1$. Since the
dependency on $\eta$ in \eqn{eq_e-ratio} is very weak in the range
$20\lesssim \eta\lesssim 100$ we plotted the ratio only for $\eta =
50$. The figure clearly shows that the stars are bound much stronger
in the potential of the binary than that of the cluster. The ratio
increases with $\lambda$, but only for very large $\lambda$ the
energies become comparable, i.e. $\lambda\approx 500$ and
$\lambda\approx 5\times10^9$ for $q=1$ and $0.1$, respectively. Such
large clusters result in masses of about $2000\mu$ and $2\times
10^{10}\mu$ for $q=1$ and $0.1$, respectively, which are
unrealistically large. Thus, even for major mergers it is justified to
neglect the potential of the cluster with an index $\gamma =2$ in
order to compute the ejected mass that allows the \BH{}s to merge.
\begin{figure}
\begin{center}
\includegraphics[width=\FW]{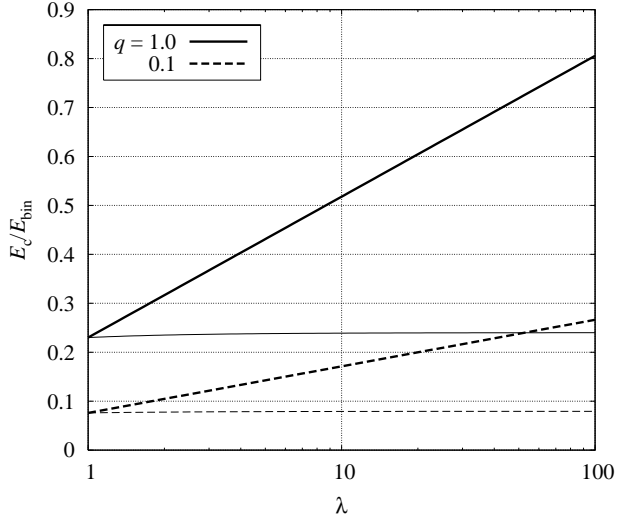}
\caption[]{The energies of the cluster in the potential of the cluster
  itself and the binary as a function of the relative cluster size
  $\lambda = r_\mrm{c}/a_\mrm{h}$. The bold and thin lines are for
  flat ($\gamma =2$) and steep ($\gamma = 3$) profiles,
  respectively. ($k=1$, $\eta=50$.)}
\label{f_e-ratio} 
\end{center}
\end{figure}

In \citetalias{zier06} we showed that while the binary probably does not decay
into the third phase if the density profile is as flat as $\gamma =2$, it will
enter the final phase for steeper profiles. Repeating the above analysis for a
power law with the index $\gamma = 3$ we obtain for the cluster potential
\begin{equation}
\Phi_\mrm{c}(r) = - \frac{GM_\mrm{c}}{r} \begin{cases} \frac{\DS r}{\DS
  \ln(r_\mrm{c}/r_\mrm{i})} \left( \frac{\DS 1}{\DS r_\mrm{i}} - \frac{\DS 1}{\DS
  r_\mrm{c}} \right), & 0\leq r\leq r_\mrm{i}\\[2.5ex] \frac{\DS 1}{\DS
  \ln(r_\mrm{c}/r_\mrm{i})} \left(1-\frac{\DS r}{\DS r_\mrm{c}} +\ln\frac{\DS
  r}{\DS r_\mrm{i}}\right), & r_\mrm{i}\leq r \leq r_\mrm{c}\\[2.5ex] 1, &
  r_\mrm{c}\leq r.
\end{cases}
\label{eq_pot-3}
\end{equation}
and hence for the energy of the cluster in its own potential in the range from
$a_\mrm{g}$ to $a_\mrm{h}$
\begin{equation}
E_\mrm{c} = \frac{G M_\mrm{c}^2}{r_\mrm{c}}\, \frac{1}{(\ln\eta)^2}\, \left(
1 -\eta + \ln\eta \right).
\end{equation}
In the same range the energy of the cluster in the potential of the binary is
(Eq.~(11) of \citetalias{zier06} multiplied with $2$):
\begin{equation}
E_\mrm{bin} = -\frac{GM_{12}M_\mrm{c}}{\ln(r_\mrm{c}/r_\mrm{i})} \left(
\frac{1}{a_\mrm{g}} - \frac{1}{a_\mrm{h}} \right) =
-\frac{GM_{12}M_\mrm{c}}{a_\mrm{h}} \frac{\eta -1}{\ln\zeta}.
\label{eq_e-b3}
\end{equation}
Assuming as before that the mass $m_\mrm{ej}$ is distributed in the cluster
between $a_\mrm{g}$ and $a_\mrm{h}$ we obtain from Eqs.~(\ref{eq_m}) and
(\ref{eq_mej})
\begin{equation}
M_\mrm{c} = \frac{\mu}{k} \ln\zeta
\frac{\ln\eta}{\ln(\zeta/\lambda)}.
\end{equation}
For $r_\mrm{i} = a_\mrm{g}$ (i.e. $\zeta=\lambda\eta$) the ratio of
the energies is
\begin{equation}
\frac{E_\mrm{c}}{E_\mrm{bin}} = \frac{1}{2k}\, \frac{q}{(1+q)^2} \left[2 -
\frac{(\lambda+1)\ln\eta}{\lambda (\eta -1)}\right].
\label{eq_e-ratio3}
\end{equation}
This ratio is displayed in \fgr{f_e-ratio} by the thin lines. We can
see that for the steeper profile the cluster potential contributes an
even smaller fraction to the potential energy of the stars and
increases much less with $\lambda$ (i.e. is almost constant) than in
the case of a shallower profile (bold lines). The steeper the density
distribution, the stronger the stars are bound to the \BH{}s. For a
density distribution with a power law index as steep as $\gamma =3$
the contribution of the self energy of the cluster is negligible, even
in the limit of large ratios $\lambda = r_\mrm{c} / a_\mrm{h}$. The
term in the square brackets of \eqn{eq_e-ratio3} then tends to $2 -
\ln(\eta)/(\eta -1)$, having a maximum of 2 if $\eta$ tends to
infinity so that the ratio $E_\mrm{c}/E_\mrm{bin}$ approaches a
maximum of $q/(1+q)^2$, what is $0.25$ and only $0.083$ for $q=1$ and
$0.1$, respectively. While neglecting the cluster's potential is well
justified for density profiles with $\gamma=2$ it is an even better
approximation for steeper mass distributions. Thus the results we
obtained in \citetalias{zier06} should be a reasonably good
approximation in the limit that the cluster contains a mass within the
radius $a_\mrm{h}$ which corresponds to about the binary's
mass. Therefore we can be confident in our results which predict a
successful merger of the \BH{}s after the ejection of about
$m_\mrm{ej}$.




\section{Mass and density profile}
\label{s_density}

The observed profiles of early type galaxies have been published in
various papers. \citet{lauer95} casted the surface brightness
distributions into the `Nuker-law', i.e. two power laws with inner
and outer slopes which match at the break radius $R_b$. This is
typically found on scales of some tens of parsecs or more. They
detected a bimodal distribution and classified sources with slopes in
the range $0\leq\beta\lesssim0.3$ as core galaxies while they referred
to steep profiles with $\beta\gtrsim0.5$ as power law
galaxies. Deprojecting the surface brightness distributions
\citet{gebhardt96} found the slopes of the luminosity density profiles
to be in the range $0.3\lesssim\gamma\lesssim 2.5$ peaking at $0.8$
(core galaxies) and $1.9$ (power law galaxies). Because the outer
slope of both types is less than $3$ it has to steepen again at large
radii to keep the mass finite. This is not covered by the
Nuker-law. Both classes of galaxies also differ in other respects
\citep{lauer95}: Core galaxies have larger cores and are more massive
and luminous with boxy or elliptical isophotes. They show a high
velocity dispersion while they are slowly rotating. This is in
agreement with this type of galaxies having undergone a major merger
with redistribution of matter, dissipation of energy and cancellation
of large amounts of angular momentum. This probably results in an
increased central density at the time the binary becomes hard, as we
argued in \citetalias{zier06}. Later
\citet{carollo97,rest01,ravindranath01} found galaxies with
intermediate slopes what might suggest that there is a smooth and
continuous variation in the slopes and other properties of the
galaxies.

More recently \cite{graham03} suggested a combination of a power law
and S\'{e}rsic law to fit the surface brightness profiles. The new
`Core-S\'{e}rsic' model seems to generate better fits to the
observed distribution, including the outer regions, and thus keeps the
total mass finite. According to this model the galaxies previously
classified as `power laws' can be fitted by a pure S\'{e}rsic
profile, while core galaxies follow a flat power law in the inner
regions and a S\'{e}rsic profile at larger distances
\citep{trujillo04}. The slopes of the inner power law are
$20\textrm{--}40\%$ larger compared to the Nuker-fits, while the break
radii are smaller by a factor of $\sim2\textrm{--}5$.  Thus the
distribution of the slopes of the spatial profiles of the core
galaxies peaks at about $1.14$. Assuming a constant mass to light
ratio we can use these models and their slopes for the density
profiles.


\subsection{Ejected mass \& cluster mass}
\label{s_mej-mc}
If the mass $m_\mrm{ej}$ is distributed between $a_\mrm{g}$ and
$a_\mrm{h}$ according to a power law with index $\gamma =3$ its
ejection allows the \BH{}s to shrink to the final separation
$a_\mrm{f} = a_\mrm{g}$ \citepalias{zier06}. For flatter profiles we
find $a_\mrm{f} > a_\mrm{g}$ while for steeper ones the \BH{}s shrink
into the range of phase 3, see Fig.~1 in that article. We also
calculated how much mass $m_\mrm{rq}$ exactly is required to be
distributed between $a_\mrm{g}$ and $a_\mrm{h}$ for other slopes than
$\gamma =3$ in order to allow the binary to enter the final
phase. This mass, expressed as a fraction of $m_\mrm{ej}$, is
\begin{equation}
\frac{m_\mrm{rq}}{m_\mrm{ej}} = \frac{2-\gamma}{3-\gamma}\,
\frac{1-\eta^{\gamma -3}}{1-\eta^{\gamma -2}}\, \frac{\eta -1}{\ln\eta}
\label{eq_m-rq}
\end{equation}
such that for $\gamma = 3$ the ratio is equal to $1$. If the amount of mass
which is distributed in the cusp region, i.e. between $a_\mrm{g}$ and
$a_\mrm{h}$, is fixed the evolution of the merger in this phase and hence
$a_\mrm{f}$ does not depend on the extension of the cluster. However, changing
the parameter $\lambda = r_\mrm{c}/a_\mrm{h}$ will influence the total cluster
mass. Scaled to the mass which is contained in the cusp, $M(a_\mrm{h})$, we
can express it as
\begin{equation}
\frac{M_\mrm{c}}{M(a_\mrm{h})} = \frac{(\lambda\eta)^{3-\gamma}
  -1}{\eta^{3-\gamma} -1},
\label{eq_m-ratio}
\end{equation}
assuming that the density distribution at larger radii has the same
slope as in the cusp region. Deriving this mass ratio from \eqn{eq_m}
we again assumed that $r_\mrm{i}=a_\mrm{g}$ so that $\zeta$ can be
replaced with $\lambda\eta$. If we require $M(a_\mrm{h})=m_\mrm{rq}$
we can express the resulting cluster mass with the help of
\eqn{eq_m-rq} in units of $m_\mrm{ej}$. This is plotted as function of
the index $\gamma$ in \fgr{f_m-ratio} (bold lines, left-hand
$y$-axis). With $m_\mrm{ej}$ being of the order of $10^{8}\, M_\odot$
for a binary of a comparable mass (see \eqn{eq_mej}), \fgr{f_m-ratio}
shows that for the \BH{}s to merge the cluster has to be of about the
same mass in case of steep profiles. The cluster mass is increasing
with decreasing $\gamma$, the more steep, the larger the cluster
is. It amounts to about $10^{10}\,M_\odot$ if $\gamma \approx 1.25$,
$1.6$ or $2.1$ for $\lambda = 5$, $10$ or $50$, respectively. Thus for
large clusters with flat density profiles, as seen in core galaxies
which are thought to have already undergone a major merger, the mass
quickly becomes unphysically large, i.e. the binary would stall. But
if the distribution is steep enough as we assume for an ongoing merger
($\gamma \gtrsim 2$) this problem does not occur.
\begin{figure}
\begin{center}
\includegraphics[width=\FW]{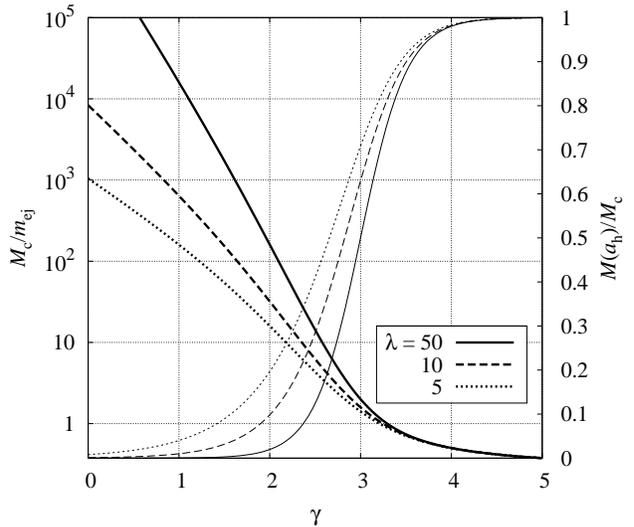}
\caption[]{Bold lines, left-hand $y$-axis: The cluster mass in units
  of $m_\mrm{ej}$ is shown under the condition that the mass within
  $a_\mrm{h}$ is large enough to allow the binary to merge. Thin
  lines, right-hand $y$-axis: fraction of the cluster mass within
  $a_\mrm{h}$.}
\label{f_m-ratio} 
\end{center}
\end{figure}

Keeping the cluster mass fixed to $m_\mrm{ej}$ and plotting the
inverse of \eqn{eq_m-ratio} we obtain the fraction of $m_\mrm{ej}$
that is distributed within $a_\mrm{h}$ and therefore accessible for
ejection by the binary (thin curves in \fgr{f_m-ratio}, scaled on the
right-hand $y$-axis). We get the same curves if we restrict the cluster
mass to $m_\mrm{rq}$ instead of $m_\mrm{ej}$ and then plot the cusp
mass within $a_\mrm{h}$ in units of $m_\mrm{rq}$. The mass fraction of
the cusp is increasing with the slope. For a small $\gamma$ the decay
of the binary stops before the transition to the third phase is
reached because there is not enough mass available for ejection (see
also Fig.~1 of \citetalias{zier06}). The larger $\lambda$ is, the less
mass is contained within this range and the earlier the binary
stalls. For steeper distributions with $\gamma \gtrsim 2$ the fraction
of ejected mass increases steeply with $\gamma$. Even for a cluster as
large as $\lambda = 50$ this mass is sufficient for $\gamma \approx 3$
to allow the binary to shrink to a radius as small as about
$2\,a_\mrm{g}$ (Eq.~(12) of \citetalias{zier06}).

Comparing the ejected with the cluster mass we come to the same
conclusion as before. For large clusters with shallow profiles its
mass becomes unphysically large if the binary is supposed to enter the
third phase. Therefore the decay of the binary would stall in core
galaxies, which probably have already undergone a major merger. This
is in good agreement with the conclusions drawn by \citet{roos81} from
his numerical three-body experiments. However, for steep profiles as
we expect them to be formed during mergers, there is sufficient mass
available for ejection without the cluster mass becoming too large, so
that the \BH{}s will coalesce. The post-merger profile has to match
the observed mass distributions and comparing both we hope to obtain
more information about the properties of the cusp.


\subsection{Mass distribution during and after the merger}
\label{s_mass-dist}
In \eqn{eq_da} we can express the infinitesimal mass in terms of the density,
$\D m = 4\pi r^2 \rho(r) \D r$, and rewrite this equation in the form
\begin{equation}
\frac{\D m}{\D r} = -\frac{\mu}{kr} = -4\pi r^2 \rho_\mrm{ej}(r).
\label{eq_dmdr}
\end{equation}
Solving for the density distribution of the cusp we obtain
\begin{equation}
\rho_\mrm{ej}(r) = \frac{\mu}{4\pi k} r^{-3},
\label{eq_rho}
\end{equation}
i.e. a profile with a power law index $\gamma =3$. This distribution
represents the solution in Fig.~1 of \citetalias{zier06}, where for $\gamma =
3$ and $\lambda = 1$ all lines go through $a_\mrm{f}/a_\mrm{g} = 1$. This
means that $m_\mrm{ej}$ is distributed with just the right steepness that
after its ejection the binary has shrunk to the semimajor axis $a_\mrm{g}$
where gravitational radiation starts to dominate the further decay. With the
slope $\gamma = 3$ the \BH{}s eject the mass $\D m$ in the distance $r$ owing
to which the binary shrinks by an amount $\D r$ that corresponds precisely to
the thickness of the spherical shell containing $\D m$. If the density would
fall below that of \eqn{eq_rho} somewhere in the range between $a_\mrm{g}$ and
$a_\mrm{h}$ and we ignore the mass outside the shell $4\pi r^2\D r$, the
binary would stall at this distance. For a steeper distribution more matter is
bound deeper in the potential at smaller radii and its ejection would allow
the binary to decay to radii smaller than $a_\mrm{g}$. If we assume that
\eqn{eq_rho} is the initial profile, all the mass becomes ejected for
$r\lesssim a_\mrm{h}$ and a hole remains in this range after the merger. But
if the initial profile exceeds that of \eqn{eq_rho} we could substract the
latter from the former to compute the density distribution after the \BH{}s
have merged.

\subsubsection*{Non-core profiles}
As \citet{trujillo04} showed, a S\'{e}rsic model is a very good
approximation to surface brightness profiles, and assuming a constant
mass to light ratio we can write for the surface density:
\begin{equation}
\label{eq_s-dens}
\Sigma(R) = \Sigma_0
\exp\left[-b\left(\frac{R}{R_\mrm{e}}\right)^{1/n}\right],
\end{equation}
where $R$ is the 2-dimensional radius and $\Sigma_0$, $R_\mrm{e}$ and
$n$ are parameters. While $\Sigma_0$ is just a scaling factor we can
relate $R_\mrm{e}$ to $n$ if we require that a disc of radius
$R_\mrm{e}$ contains half of the total mass. Integrating
\eqn{eq_s-dens} from the center to $R_\mrm{e}$ and infinity the
comparision of the obtained masses yields the relation $\Gamma(2n) =
2\Gamma(2n, b)$. $\Gamma$ with one argument denotes the gamma function
and with two arguments the incomplete gamma function, where we used
the notation of \citet{abramowitz72}. A good approximation to this
relation is given by \citet{prugniel97} with $b=2n -1/3 +
0.009876/n$. They also deprojected the surface density profile and
gave a simplified fit to the spatial density distribution which can be
written as
\begin{equation}
\label{eq_rho-dp}
\rho_\mrm{gal}(r) = \rho_0 \left(\frac{r}{R_\mrm{e}}\right)^{-p}
\exp\left[-b\left(\frac{r}{R_\mrm{e}}\right)^{1/n}\right].
\end{equation}
This is a S\'{e}rsic model multiplied with a power law where $\rho_0 =
\Sigma_0/R_\mrm{e}$ and $r$ is the 3-dimensional radius. The surface
integral of \eqn{eq_s-dens} should give the same total mass as the
volume integral of \eqn{eq_rho-dp} and we obtain $\Gamma(2n) = 2
b^{n(p-1)} \Gamma[n(3-p)]$. This allows to compute $p$ as function of
$n$. \citet{marquez00} give an updated version of the approximation by
\citet{lima99}, $p=1.0 - 0.6097/n + 0.05563/n^2$. Note that the half
mass radius of the spatial distribution is slightly larger than of the
surface profile, $r_\mrm{e}/R_\mrm{e} = 1.356 - 0.0293/n + 0.0023/n^2$
(\citeauthor{lima99}).

Non-core galaxies do not show signs of a recent merger like core
galaxies. On the other hand the innermost data points used for fits
are larger than $a_\mrm{h}$, and therefore a transition at this
distance to a possible inner power law could not have been
resolved. By the time the binary becomes hard we assume the mass to be
distributed according to one of the following models: (a) According to
the deprojected S\'{e}rsic model, which describes the resolved part
very good and continues farther down to the center. After the \BH{}s
have merged we have to substract the ejected cusp from the deprojected
S\'{e}rsic profile, which therefore has to be at least as large as
that of the ejected cusp at $r=a_\mrm{g}$. In model (b) the mass
distribution is that of a deprojected S\'{e}rsic model with an
additional cusp in the center when $a=a_\mrm{h}$, leaving behind a
deprojected S\'{e}rsic profile after the \BH{}s have ejected the
cusp. The distribution of the ejected mass fraction in the cusp region
for both models is:
\begin{equation}
\label{eq_rho-gamma}
\rho_\gamma = \rho_\mrm{h} \left(\frac{r}{a_\mrm{h}}\right)^{-\gamma}
\end{equation}
It has to match \eqn{eq_rho-dp} at $r=a_\mrm{g}$ and $a_\mrm{h}$ for
model (a) and (b), respectively. The volume integration of
\eqn{eq_rho-gamma} over the cusp range yields the ejected mass
fraction $M_\mrm{ej}$.

Equating both mass distributions at $r=a_\mrm{g}$ we obtain for the
galaxy profile of model (a):
\begin{equation}
\label{eq_rho-a}
\rho_\mrm{gal,(a)} = \rho_\mrm{h} \frac{\eta^{\gamma -p}}{x^p} \exp
\left\{ -b\left[\left(\frac{x}{x_\mrm{e}}\right)^{\frac{1}{n}} -
  \left(\frac{1}{\eta x_\mrm{e}}\right)^{\frac{1}{n}} \right]\right\},
\end{equation}
with $x = r/a_\mrm{h}$ and $x_\mrm{e} =
R_\mrm{e}/a_\mrm{h}$. $\rho_\mrm{h}$ is obtained from \eqn{eq_rho0}
using $r_0 = r_\mrm{c} = a_\mrm{h}$, $r_\mrm{i} = a_\mrm{g}$ and
$M_\mrm{c} = M_\mrm{ej}$. We denote the term on the right hand side of
the curly bracket of \eqn{eq_rho0} with $g(\gamma, \eta)$ and thus
find $\rho_\mrm{h} = g(\gamma, \eta)\,M_\mrm{ej}/(4\pi
a_\mrm{h}^3)$. The volume integration then yields for the total mass
of the galaxy in units of the ejected mass
\begin{equation}
\label{eq_mg-a}
\frac{M_\mrm{gal,(a)}}{M_\mrm{ej}} = \eta^{\gamma -p}\frac{n\,
  x_\mrm{e}^{3-p}}{b^{n(3-p)}}\, g(\gamma,\eta)\, \Gamma[n(3-p)]\,
  e^{b(1/\eta x_\mrm{e})^{1/n}}
\end{equation}
For model (b) with the condition $\rho_\gamma (a_\mrm{h})
=\rho_\mrm{gal} (a_\mrm{h})$ we obtain for the density and the mass:
\begin{figure}
\begin{center}
\includegraphics[width=\FW]{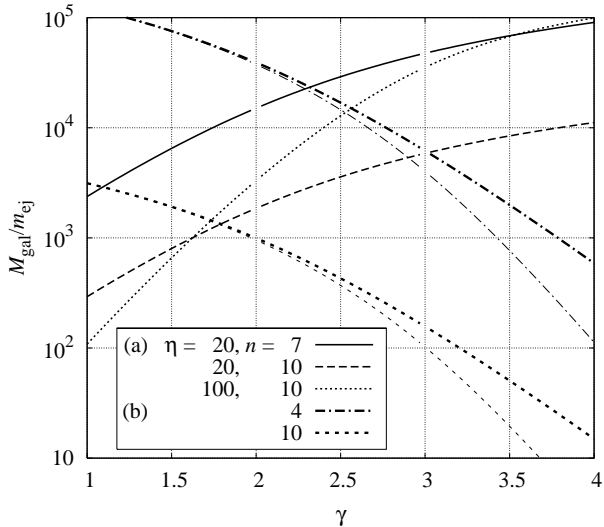}
\caption[]{Mass of a non-core galaxy as function of the slope $\gamma$
  of the ejected profile. $x_\mrm{e}=10^3$ and $m_\mrm{ej}\approx
  10^8\, M_\odot$. The bold and thin lines in model (b) are for $\eta
  = 20$ and $100$, respectively.}
\label{f_m-gal} 
\end{center}
\end{figure}
\begin{align}
\label{eq_rho-b}
\rho_\mrm{gal,(b)} &= \rho_\mrm{h} \frac{1}{x^p} \exp
\left\{ -b\left[\left(\frac{x}{x_\mrm{e}}\right)^{\frac{1}{n}} -
\left(\frac{1}{x_\mrm{e}}\right)^{\frac{1}{n}} \right]\right\}\\
\intertext{and}
\label{eq_mg-b}
\frac{M_\mrm{gal,(b)}}{M_\mrm{cusp}} &= \frac{n\,
  x_\mrm{e}^{3-p}}{b^{n(3-p)}}\, g(\gamma,\eta)\, \Gamma[n(3-p)]\,
  e^{b(1/x_\mrm{e})^{1/n}}.
\end{align}
While the density given in \eqn{eq_rho-a} refers to the initial
profile from which we have to substract the cusp of \eqn{eq_rho-gamma}
to obtain the final distribution, \eqn{eq_rho-b} is the expression for
the final profile after the ejection of $\rho_\gamma$. Because the
ejected mass is only a minor fraction of the total galaxy mass it does
not make a difference whether we compare the initial or final galaxy
masses. Note that they are actually lower limits for a successful
merger. Assuming that $M_\mrm{ej}=m_\mrm{rq}$ of \eqn{eq_m-rq} we plot
the galaxy mass in units of $m_\mrm{ej}$ as function of the inner
slope $\gamma$ in \fgr{f_m-gal}. For all plots we used a ratio
$R_\mrm{e}/a_\mrm{h} = 10^3$ because \citet{trujillo04} obtained for
the fits of non-core galaxies half-light radii typically on the kpc
scale. While we obtain ascending curves (solid, dashed, dotted) for
model (a), they are descending (short dashed, dash-dotted) in case
of model (b). If we keep the ejected mass of the cusp constant between
$a_\mrm{g}$ and $a_\mrm{h}$ but increase the slope $\gamma$ of its
distribution the density increases at the inner edge, being
compensated for by a decrease at the outer edge. Because the densities
of model (a) and (b) match the cusp at $a_\mrm{g}$ and $a_\mrm{h}$,
respectively, we obtain the different dependencies on $\gamma$. For
galaxies with a mass of some $10^{12}\,M_\odot$ or less \fgr{f_m-gal}
shows that possible solutions for model (a) with $\gamma\gtrsim 2$
require a large shape parameter $n\gtrsim 7$. We also find small cusps
($\eta\approx 20$) to be more likely than larger ones. If on the other
hand an additional cusp is formed in the center when $a=a_\mrm{h}$ we
do not find tight restrictions for the parameters. In the range
$\gamma\gtrsim 2.5$ all shape parameters $n\gtrsim 4$ are
possible. The size of the cusp $\eta$ ($20$ and $100$ for bold and
thin lines, respectively) has only a small influence on the total mass
of the galaxy. \citet{trujillo04} obtained for their best fits to
non-core galaxies an average shape parameter of about $5$. This
basically excludes model (a) and is in very good agreement with steep
cusps ($\gamma\gtrsim 2.5$) in model (b) so that shallower profiles
seem to be less likely.
\begin{figure}
\begin{center}
\includegraphics[width=\FW]{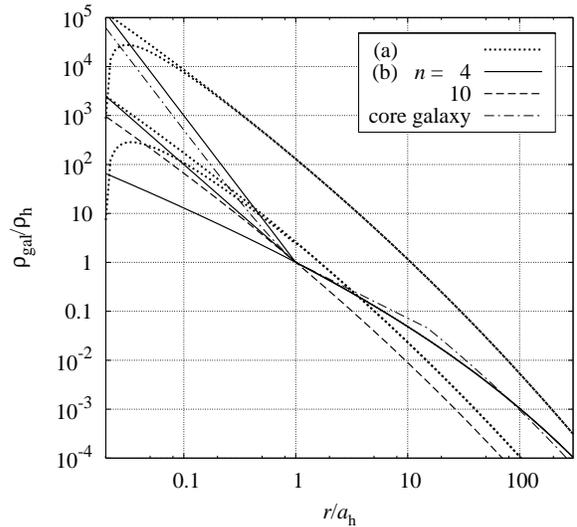}
\caption[]{Density profiles. Model (a): The dotted lines with
  parameters $\gamma =3$ and $2$ (upper and lower pair,
  respectively). Initial and final profiles are the upper and lower
  branch of each pair. $n=10, \eta=50$. Model (b): solid lines from
  top to bottom: initial profiles with $\gamma=3$, $2$ and final
  profile for $n=4$. Dashed line: final profile with $n=4$. A core
  profile is shown by the thin dash-dotted line with $n=6$.}
\label{f_rho-nc} 
\end{center}
\end{figure}

In \fgr{f_rho-nc}, we plotted the density profiles for both models. The
upper branch of both dotted pairs of curves shows the initial profile
of model (a), i.e. a deprojected S\'{e}rsic model distribution with
$n=10$ and $\eta =20$. After the \BH{}s have coalesced and stars have
been ejected from the center we obtain the lower branches as final
profiles. For the upper and lower pair we used $\gamma=3$ and $2$,
respectively, for the ejected distribution. In the latter case we
multiplied the final profile in the region $r<a_\mrm{h}$ with a factor
of $1.5$ in order to avoid a kink at $r=a_\mrm{h}$. This does not
matter because \fgr{f_rho-nc} is meant to only give a rough idea about
the possible shape of the profiles. Model (b): Both solid ($n=4$)
upper branches depict the initial profile for $\gamma=3$ (top) and
$2$. The lower branch is the final profile which is a pure deprojected
S\'{e}rsic model and extends to the outer regions. The dashed line
shows the final distribution for $n=10$, which has a slope almost as
steep as $2$. The ratio of the inital and final mass contained in the
cusp region varies between $1.8$ and $3.3$ for the parameters $(\eta,
n)= (20, 10)$ and $(100, 4)$, respectively, if $\gamma = 2.5$. For
$\gamma =3$ we obtain a ratio between the limits $3.5$ and $8.5$ for
the same sets of parameters. Hence this amount of mass has to be added
to the mass of the relaxed profile in the cusp region during the
merger if the \BH{}s are to coalesce.

Our results suggest that if non-core galaxies are merger remnants then
model (b) is more likely than (a) and a cusp has been formed during
the merger with an index $\gamma\gtrsim 2.5$ when $a=a_\mrm{h}$, which
is subsequntly removed by the merging \BH{}s. This is also what we
expect: It would be quite unlikely that an undisturbed deprojected
S\'{e}rsic profile is maintained during the merger, especially when
both cores merge and the binary becomes hard, without the formation of
an inner cusp.

\subsubsection*{Core profiles}
The surface brightness of core galaxies is well fitted by a S\'{e}rsic
model in the outer parts and by a flat power law at distances smaller
than the break radius $R_b$, see \citet{trujillo04}. To keep things
simple we will not try to deproject the core-S\'{e}rsic
profile. Instead we use a power law with the index $\delta$ in the
core region below the 3-dimensional break radius $r_b$, which we
identify with $R_b$, and a deprojected S\'{e}rsic profile at larger
radii, like \citet{terzic05}. For an infinitely sharp transition
between both regimes the densities in the core and outer galaxy region
can be written in the form:
\begin{align}
\rho_\mrm{core} &= \rho_b \left(\frac{x}{x_b}\right)^{-\delta}\\
\rho_\mrm{gal} &= \rho_b \left(\frac{x}{x_b}\right)^{-p} \exp\left\{
-b\left[ \left(\frac{x}{x_\mrm{e}}\right)^{\frac{1}{n}} -
  \left(\frac{x_b}{x_\mrm{e}}\right)^{\frac{1}{n}} \right] \right\}.
\end{align}
Here $x$ denotes the radius in units of $a_\mrm{h}$ so that $x_b =
r_b/a_\mrm{h}$ and $x_\mrm{e} = R_\mrm{e}/a_\mrm{h}$. Both profiles
match at $x_b$ with the density $\rho_b$. Because
\citeauthor{trujillo04} found the slope of a core-S\'{e}rsic
approximation to be about 30\% steeper than Nuker fits in the core
region we use $\delta = 1.14$ for the deprojected slopes instead of
$0.8$ what has been obtained by \citet{gebhardt96} when deprojecting
Nuker fits. For the seven core galaxies of the sample of
\citeauthor{trujillo04} the average ratio of the half light radius and
the break radius is $R_\mrm{e}/R_b\approx 70$. We assume that this
ratio is roughly maintained for the deprojected spatial radii. Because
$R_\mrm{e}$ in this sample is typically on the kpc scale so that
$x_\mrm{e}=10^3$ we find for the spatial break radius $x_b\approx 15$,
in units of $a_\mrm{h}$. For a flat core to be able to exist within
$r_b$ its slope has to be less steep than the slope $\alpha$ of the
deprojected S\'{e}rsic profile evaluated at $x=x_b$:
\begin{equation}
-\alpha\equiv \frac{\D \log\rho_\mrm{gal}}{\D \log x} = -\left[p +
  \frac{b}{n} \left(\frac{x}{x_\mrm{e}}\right)^{1/n} \right]
\end{equation}
The condition $\alpha > \delta$ is always fulfilled for $n\gtrsim 2$
and hence imposes no restrictions on our model. The break radius is
larger by a factor of about 10 compared to $a_\mrm{h}$, which is on
the parsec scale and thus very close to the radius of the innermost
data point which has been used in the fits. Hence, if the profile of
core galaxies is about the same at $r>a_\mrm{h}$ before and after the
merger, it will have an additional cusp $\rho_\gamma$ at $r<a_\mrm{h}$
when the binary becomes hard. By the time the \BH{}s have coalesced we
assume this cusp to have been removed and the profile of the core
region ($a_\mrm{h}\leq r\leq r_b$) to extend also into the cusp
region. Matching both profiles at $x=1$ we obtain $\rho_b =
\rho_\mrm{h} x_b^{-\delta}$ and find for the initial profile at
$a_\mrm{g}\leq r\leq a_\mrm{h}$:
\begin{equation}
\rho_\mrm{cusp} = \rho_\gamma + \rho_\mrm{core} = \rho_\mrm{h} \left(
x^{-\gamma} + x^{-\delta} \right).
\end{equation}
The complete mass distribution of such a core galaxy from the cusp to
the outer galaxy region is depicted by the thin dash-dotted curve in
\fgr{f_rho-nc} for the time when $a=a_\mrm{h}$. The used parameters
are $\gamma = 3$, $\delta=1.14$, $x_b = 15$, $x_\mrm{e} = 10^3$ and
$n=6$. Performing the volume integrations of all the components of the
density profile over their respective regions we obtain for the
masses:
\begin{align}
\label{eq_m-cusp-nc}
M_\mrm{cusp} &= M_\mrm{ej} \left[1 + g(\gamma, \eta) \frac{1-\eta^{\delta
      -3}}{3-\delta} \right],\\
\label{eq_m-core-nc}
M_\mrm{core} &= M_\mrm{ej}\,g(\gamma, \eta)\,\frac{x_b^{3-\delta}
-1}{3-\delta},\\
\begin{split}
\label{eq_m-gal-nc}
M_\mrm{gal} &= M_\mrm{ej}\,g(\gamma, \eta)\,
\frac{x_\mrm{e}^{3-p}}{x_b^{\delta-p}}\, \frac{n}{b^{n(3-p)}}\\
&\phantom{==} \Gamma\left[ n(3-p),
b\left(\frac{x_b}{x_\mrm{e}}\right)^{1/n}\right] \exp \left[
b\left(\frac{x_b}{x_\mrm{e}}\right)^{1/n}\right].
\end{split}
\end{align}
As before we used $\rho_\mrm{h} = g(\gamma, \eta)\,M_\mrm{ej}/(4\pi
a_\mrm{h}^3)$ with $g(\gamma, \eta)$ being the term right of the curly
bracket in \eqn{eq_rho0} with $r_0 = r_\mrm{c} = a_\mrm{h}$ and
$r_\mrm{i} = a_\mrm{g}$. For a successful merger we identify
$M_\mrm{ej}$ with $m_\mrm{rq}$ of \eqn{eq_m-rq} and plotted the total
mass of the galaxy $M_\mrm{tot} = M_\mrm{cusp} + M_\mrm{core} +
M_\mrm{gal}$ in units of $m_\mrm{ej} \approx 10^8\,M_\odot$ as
function of $\gamma$ in \fgr{f_m-cgal}. We used $x_b =15$,
$x_\mrm{e}=10^3$ and $\delta = 1.14$. The bold and thin lines are for
$\eta = 100$ and $20$, respectively, and from top to bottom we used as
shape parameter $n=4$, $6$ and $10$ for each set of curves. The figure
shows that the lower mass limit that enables the \BH{}s to merge
decreases with increasing $\gamma$. Because the cusp profile has to
match the flat core distribution at $r=a_\mrm{h}$, which in turn
matches the deprojected S\'{e}rsic model at $r=r_b$, the mass of the
core and galaxy region decreases with increasing $\gamma$. In fact,
for flat cusps with $\gamma\lesssim 2$ the lower mass limit might
actually be too large. While for $\gamma \lesssim 2.7$ small cusps
with $\eta = 20$ result in smaller limits than $\eta = 100$ we find
this reversed for steeper slopes. \citeauthor{trujillo04} found shape
parameters of $n\approx 5$ what argues for $\gamma\gtrsim 2.5$ and a
transient cusp if the total mass should not exceed $10^4\,m_\mrm{ej}$.
\begin{figure}
\begin{center}
\includegraphics[width=\FW]{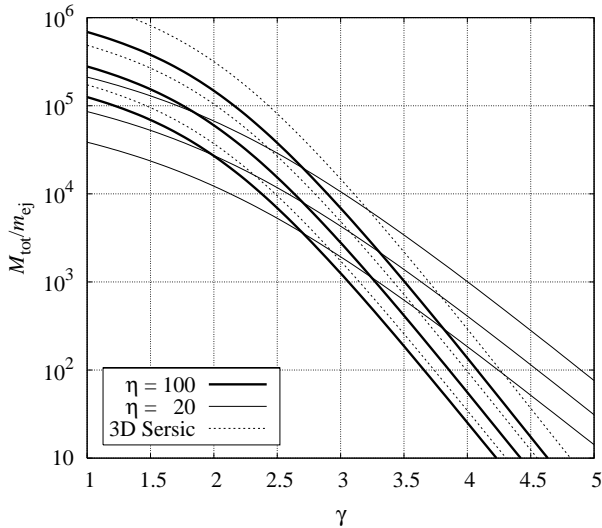}
\caption[]{Total mass $M_\mrm{tot} =
  M_\mrm{cusp}+M_\mrm{core}+M_\mrm{gal}$ as function of the slope
  $\gamma$ of the ejected profile. Fixed parameters are $(\delta, x_b,
  x_\mrm{e}) = (1.14, 15, 10^3)$. Each set of three curves is, from
  top to bottom, for the shape parameters $n=4, 6$ and $10$. The solid
  lines are for the deprojected S\'{e}rsic profile and dotted lines
  for a three-dimensional S\'{e}rsic profile with
  $\eta=100$. $m_\mrm{ej}\approx 10^8\,M_\odot$.}
\label{f_m-cgal} 
\end{center}
\end{figure}

Instead of the deprojected S\'{e}rsic model we also used the
S\'{e}rsic law of \eqn{eq_s-dens} for the spatial density distribution
at $r>r_b$. We just replace $R$ and $R_\mrm{e}$ with $r$ and
$r_\mrm{e}$, respectively, and substitute $\Sigma_0$ with $\rho_0$. If
we use a three-dimensional S\'{e}rsic model at all radii the condition
that a sphere of radius $r_\mrm{e}$ contains half of the total mass
leads to a very similar relation as in the 2-dimensional case between
the factor $b_3$ in the exponent and the shape parameter $n$,
$2\Gamma(3n, b_3) = \Gamma(3n)$. This can be approximated by $b_3 = 3n
-1/3 +0.0079/n$ \citep{merritt06}. While the cusp and the core are the
same as before we obtain for the mass in the region $r>r_b$ with the
3-dimensional S\'{e}rsic law
\begin{equation}
\frac{M_\mrm{gal,s}}{M_\mrm{ej}} = g(\gamma, \eta)\,
\frac{x_\mrm{e}^{3}}{x_b^{\delta}}\, \frac{n}{b_3^{3n}} \Gamma\left[ 3n,
b_3\left(\frac{x_b}{x_\mrm{e}}\right)^{1/n}\right]
e^{b_3(x_b/x_\mrm{e})^{1/n}}.
\end{equation}
This expression is also obtained by simply setting $p=0$ and replacing
$b$ with $b_3$ in \eqn{eq_m-gal-nc}. The lower limit for the total
mass of this model is shown by the dotted lines in \fgr{f_m-cgal} and
is typically larger by a factor of a few compared to the deprojected
S\'{e}rsic model. Hence the latter profile might be more appropriate
for the mass distribution in galaxies which is derived from surface
brightness profiles, while the former seems to describe galaxy sized
dark matter halos slightly better \citep{merritt06}. On the other hand
we can not identify the shape parameters of both models with each
other and would expect $n$ of the 3-dimensional S\'{e}rsic model to be
slightly larger than $n$ of the deprojected 2-dimensional S\'{e}rsic
profile of the surface brightness. Thus the resulting curves should be
closer to each other than shown in \fgr{f_m-cgal}.

For both, non-core and core galaxies we find in Figs.~\ref{f_m-gal}
and \ref{f_m-cgal} that for a sucessful merger generally less mass is
required for large $n$. However, on average the sources considered by
\citet{trujillo04} have a shape parameter of about $5$ which is in
favour of a steep cusp which is transiently formed (model (b) for
non-core galaxies). If core galaxies are merger remnants such a cusp
has been removed during the second phase. Constructing their profiles
with an additional cusp at the time when the binary becomes hard
yields galactic masses which are about $10^{12}\,M_\odot$ or less for
$n\approx 5$, provided that the slope of the ejected mass distribution
is $\gamma\gtrsim 2.5$. This confirms our conjecture that a steep cusp
is formed transiently in the central regions and is in favour of core
galaxies being rather the end product of a merger. However, in this
simple model we would expect the break radius in the merger remnant to
roughly coincide with $a_\mrm{h}$, the distance where the binary
becomes hard. This is not clearly defined and $a_\mrm{h}$ might
actually be larger than assumed here. Maybe before the binary becomes
hard the inner parts of the cluster surrounding each \BH are so deeply
bound in its potential that they increase the effective mass of the
\BH{}s. That would result in sligshot ejection of stars at larger
radii before the naked \BH{}s form a hard binary, i.e. a shift of
$a_\mrm{h}$ to larger distances.  Increasing the size of the cusp at
the expense of the core region, i.e. increasing $a_\mrm{h}$ while
keeping $r_b$ constant, would shift the matching point of both
profiles to larger radii where the steeper cusp profile has decreased
to smaller values, see the core profile in
\fgr{f_rho-nc}. Consequently the mass of the matching core and galaxy
region would decrease below the values shown in \fgr{f_m-cgal}, so
that $M_\mrm{tot}$ becomes $\approx 10^4\,m_\mrm{ej}$ even for
$\gamma\approx 2$. It is also possible that the difference between
$a_\mrm{h}$ and $r_b$, of about a factor of 10, is caused by
relaxation processes. We would expect a merger which shifts matter
from the region inside $a_\mrm{h}$ to larger distances also to change
the steepness of the profile as well as to smooth out sharp
transitions. In the post-merger profile the former transition between
the ejected and remaining distribution might be much closer to the
observed $r_b$ than $a_\mrm{h}$. Eccentric orbits will also result in
larger $a_\mrm{h}$, see the next Section.

In some galaxies the inner density profile has been observed to
actually decrease towards smaller radii \citep{lauer02}. Depending on
how close the ejected profile approaches the initial distribution all
slopes less then the initial one, even holes in the profile, are
possible in the merger remnant. The density distributions shown by
\citeauthor{lauer02} do not resemble a power law in the inner parts
and are more similar to the profiles resulting from simulations of a
fixed binary embedded in a stellar cluster
\citep[Fig.~1,][]{zier01}. Although these profiles would be in
agreement with a stalled binary they could also be generated by
completely merged \BH{}s, depending on the initial distribution.

\subsection{Possible deviations from the profile}
According to our analysis the profile fulfills the following
conditions when the binary becomes hard: It allows the \BH{}s to merge
completely without the cusp and galaxy becoming too massive and is
generally in agreement with observed post-merger profiles after a
fraction of the cusp has been ejected by the binary. In the cusp
region the profile is steep ($\gamma\gtrsim2.5$) turning into a power
law at $a_\mrm{h}$ with index $\delta\approx 1$ before it becomes a
deprojected S\'{e}rsic profile at $r_b\approx 10\, a_\mrm{h}$.

If the orbits of the stars in the cluster would be eccentric instead
of circular, as we assumed, and we keep the pericenter fixed, the
kick-parameter would be smaller (\eqn{eq_de-star}). This would result
in a larger mass which has to be ejected for coalescence
(\eqn{eq_mej}), but also in a less steep cusp. Stars with apocenters
$r_+ > a_\mrm{h}$ and pericenters $r_-\lesssim a_\mrm{h}$ will
probably be close to $r_+$, where they spend most of the time of the
orbital period, when the binary becomes hard. Half a period later they
interact with the \BH{}s and can become ejected. These stars could
compensate for the additional mass which is required for a successful
merger for cusps with $\gamma < 3$. The ejection of these stars shifts
$a_\mrm{h}$ to larger radii and helps to explain the difference
between $a_\mrm{h}$ and the break radius because their apocenters
populate the region between $a_\mrm{h}$ and $r_b$. For apocenters $r_+
> r_b$ the eccentricity would become larger than $(r_b -
a_\mrm{h})/(r_b + a_\mrm{h}) \approx 0.8$, what might be
unlikely. However, the exact shape of the initial profile will depend
on the mass and velocity distributions in the isolated galaxies prior
to their collision as well as on the orientation of both galactic
spins and their orbital angular momentum relative to each other. The
amount of dissipated energy, cancelled components of angular momentum
and mass with low angular momentum which is funelled into the central
regions depend on these parameters. Each of the \BH{}s will carry a
stellar cusp of about its own mass. By the time the \BH{}s become hard
both cups will have merged and together with other matter that has
accumulated in the center they form a new massive and steep
cusp. Fig.~3 of \citet{milos01} suggests that the slope of this cusp
is substantially steeper than $\gamma =2$. However, it needs to be
clarified whether this is physical or due to spurious numerical
relaxation. The post-merger profile will also be influenced by the
time scale on which the binary decays in the second phase
(Section~\ref{s_shrinking-rate}). This will depend on the eccentricity
of the \BHB and also on the initial distribution of the mass and
velocity of the stars in the cusp.

Our derivation above for the density distribution might be
oversimplified because of the following possible deviations: A
fraction of stars which has been ejected from the region $r\le
a_\mrm{h}$ will stay bound to the cluster, increasing the density at
larger radii (Section~\ref{s_k-potential}). The binary also heats the
remaining stellar population at $r\le a_\mrm{h}$, further diminishing
the cluster's density in this range. Due to the mass transfer from the
inner to the outer regions of the cluster the stars which remain in
the center are not as tightly bound as before and consequently extend
to larger radii. All this redistribution of mass leads to a profile
which is shallower than the difference between the initial and the
ejected mass distributions. However, computing the difference between
the core-S\'{e}rsic model and the S\'{e}rsic part extrapolated into
the core region, \citet{graham04} obtains mass deficits between $1$
and $2$ times $M_{12}$, depending on which method is used to determine
the post-merger \BH mass. Using Eqs.~(\ref{eq_mej}) and
(\ref{eq_m-rq}) we find for the ejected mass which allows the binary
to enter the final phase:
\begin{equation}
\label{eq_m-def}
\frac{m_\mrm{rq}}{M_{12}} = \frac{q}{(1+q)^2}\, \frac{2-\gamma}{3-\gamma}\,
\frac{1-\eta^{\gamma -3}}{1-\eta^{\gamma -2}} (\eta -1).
\end{equation}
This ratio is displayed in \fgr{f_m-def} as function of $q$ with the
slope $\gamma$ and the cusp size $\eta$ as parameters. For a single
major merger of a galaxy this is in very good agreement with the
results of \citeauthor{graham04}. Deriving \eqn{eq_m-def} we assumed
efficient energy extraction via slingshot ejection (i.e. $\epsilon =0$
and $k=1$) and therefore consider the ratios in \fgr{f_m-def} as lower
limits. Hence for mass ratios between $1$ and $2$ the cusp seems to be
steeper than $\gamma=2$ with $\gamma\gtrsim 2.5$ being more likely and
thus strengthening our arguments.
\begin{figure}
\begin{center}
\includegraphics[width=\FW]{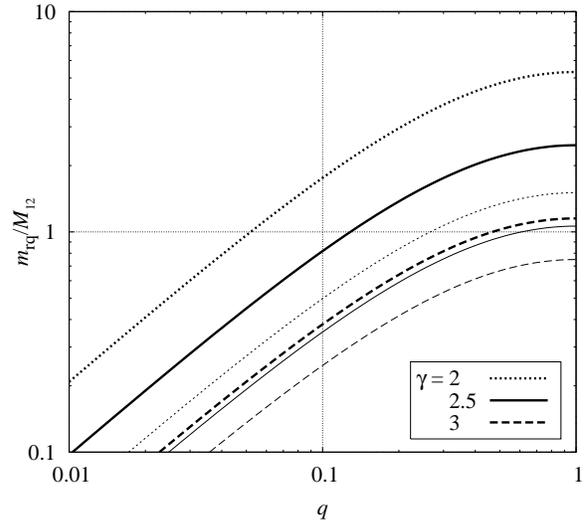}
\caption[]{Ejected mass required for a successful merger is displayed
  in units of the mass of the binary as function of the mass ratio
  $q=m_2/m_1$. The bold lines are for $\eta=100$, and the thin lines
  for $\eta=20$.}
\label{f_m-def} 
\end{center}
\end{figure}


\subsection{Evolution time scale of an isolated cusp}
\label{s_trelax}
A first approximation to the relaxation time of a collisionless system
of particles is $t_\mrm{relax}\approx t_\mrm{cross}\,0.1 N/\ln N$
\citep{binney94}, where $t_\mrm{cross}$ is the crossing time and $N$
the number of stars in the system. For our cluster with $10^8$ stars,
a size of about $1\,\mrm{pc}$ and a typical velocity of
$v_\mrm{typ}\approx\sqrt{GNm_\ast/a_\mrm{h}}$ we obtain a relaxation
time of about $10^9\,\mrm{yr}$. However, the derivation of the
expression of the relaxation time assumes a constant typical velocity
for the cluster, which requires a density profile $\propto r^{-2}$. On
the other hand the surface density is assumed to be constant as well,
which is obtained for $\rho\propto r^{-1}$ for a distribution
extending to $r=0$. For a cusp with a profile as steep as $\gamma = 3$
the relaxation time will be shorter in the inner than in the outer
regions and will also be affected by the boundary conditions.

A more precise expression for $t_\mrm{relax}$, which also takes into account
encounters between stars, can be derived with the help of the diffusion
coefficients in the Fokker-Planck approximation. We call this the diffusion
time and following \cite{binney94} we obtain
\begin{equation}
\label{eq_t-diff}
t_\mrm{diff} = \frac{\sqrt{6}}{4\pi}\, \frac{v^3(r)}{G^2 m_\ast \rho(r)
  \ln\Lambda},
\end{equation}
with $\Lambda = b_\mrm{max} v_\mrm{typ}^2/2Gm_\ast$ and $b_\mrm{max}$
being the maximum impact parameter, which is of the order of the size
of the cusp. Because only the logarithm of $\Lambda$ enters
\eqn{eq_t-diff}, even larger errors of $\Lambda$ result only in minor
errors of the diffusion time. However, the relaxation time depends
sensitively on the velocity, which is of the order of the circular
velocity $v_\mrm{c} = \sqrt{G M(r)/r}$. While for a steep cusp with
$\gamma = 3$ the density in the denominator of \eqn{eq_t-diff} will
diminish the diffusion time in the inner regions, the circular
velocity counteracts this tendency and increases with decreasing $r$
(being a factor of more than $3$ larger for $\gamma=3$ than for
$\gamma =2$ at $r\approx 3\,a_\mrm{g}$). \fgr{f_t-diff} shows
$t_\mrm{diff}$ as a function of the radius for $\gamma = 2$ and
$3$. As expected the evolution times are much smaller at small radii
for both slopes of the density distribution. Surprisingly the
relaxation times of the steeper cusp (bold dashed curve) are larger
than of the flatter cusp (bold dotted curve) over the whole plotted
range, even at smaller radii. This is due to the strong dependency on
the velocity. If we replot $t_\mrm{diff}$ and keep the velocity fixed
to $v_\mrm{typ} = \sqrt{GM_\mrm{c}/a_\mrm{h}}$ we obtain the thin
curves, which are more consistent with our expectations: In the inner
parts the steeper cusp evolves faster, but slower in the outer regions
than a flatter distribution with $\gamma = 2$. The difference between
both dotted curves, increasing towards the inner edge, is due to the
influence of the cut-off at the inner boundary of the mass
distribution on the circular velocity. However, a constant velocity
all over the cusp region does not seem to be appropriate if the slope
of the mass distribution is different from $\gamma = 2$.
\begin{figure}
\begin{center}
\includegraphics[width=\FW]{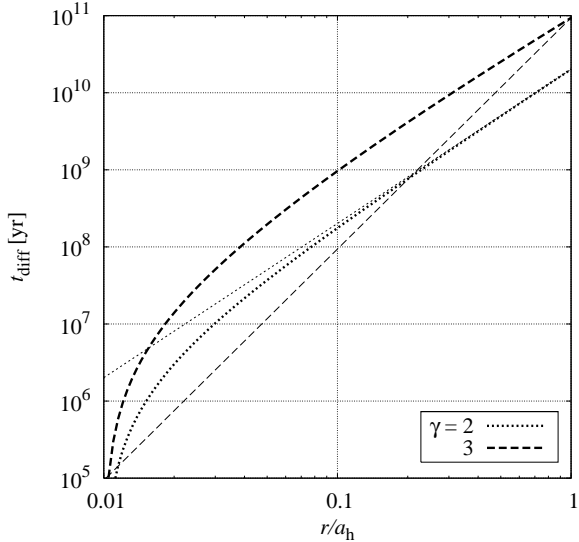}
\caption[]{Relaxation time of the cusp as function of the radius, derived
  using diffusion coefficients. Bold lines: $v_\mrm{c}^2\propto M(r)/r$, thin
  lines: $v_\mrm{c} = \mathit{const.}$}
\label{f_t-diff}
\end{center}
\end{figure}

To check this result we compare it with the time a star needs to
spiral inwards from a distance $r > a_\mrm{g}$ to $a_\mrm{g}$ due to
dynamical friction. The decelerating force $F=m_\ast\, \D v/\D t$ is
\begin{equation}
\label{eq_f-decel}
F = -\frac{8\pi G^2 m_\ast^2 \rho(r)\ln\Lambda}{v_\mrm{c}^2(r)}
\left(\text{erf}(\chi) - \frac{2\chi}{\sqrt{\pi}} e^{-\chi^2}\right),
\end{equation}
with $\text{erf}$ being the error function and $\chi\equiv
v_\mrm{c}/(\sqrt{2}\sigma)$ (\citeauthor{binney94}). This force is
tangential to the orbit and therefore causes the star to lose angular
momentum per unit mass at a rate
\begin{equation}
\label{eq_l-loss}
\frac{\D l_\ast}{\D t} = \frac{F r}{m_\ast}.
\end{equation}
The star continues to orbit at the speed $v_\mrm{c}(r)$ as it spirals
into the center with an orbital angular momentum per unit mass $l_\ast
= rv_\mrm{c}$. If $\gamma = 2$ and the cusp extends to the origin the
circular velocity is constant. The integration of \eqn{eq_l-loss}
yields for the time to spiral inwards from an initial radius
$r_\mrm{in}$ to $a_\mrm{g}$
\begin{equation}
\label{eq_t-fric0}
t_\mrm{fric} = \frac{v_\mrm{c}^3}{8\pi G^2 m_\ast S(\chi)\ln\Lambda}
\frac{x_\mrm{in}^\gamma - \eta^{-\gamma}}{\gamma \rho_0}.
\end{equation}
Here $S(\chi)$ is just an abbreviation for the term in the brackets of
\eqn{eq_f-decel}, and $x$ is the radius in units of $a_\mrm{h}$. If
the distribution has an inner boundary at $a_\mrm{g}$ the circular
velocity is not constant anymore what has to be taken into account in
the time derivative of $l_\mrm{\ast}$ in \eqn{eq_l-loss}. Then the
time to spiral to the inner boundary becomes
\begin{equation}
\begin{split}
\label{eq_t-fric2}
t_\mrm{fric} = & \frac{Q}{2\eta^2\sqrt{1-1/\eta}} \Big[\eta(2\eta
  x_\mrm{in} -3)\sqrt{x_\mrm{in}(x_\mrm{in} -1/\eta)}\\ & +
  \ln(\sqrt{\eta x_\mrm{in}} + \sqrt{\eta x_\mrm{in} -1})\Big],
\end{split}
\end{equation}
with $Q = \sqrt{N a_\mrm{h}^3/(Gm_\ast)}/(4S(\chi)\ln\Lambda) \approx
4\times 10^9\,\mrm{yr}$ for our cusp and $M_\mrm{c} = Nm_\ast$. Also
for $\gamma =3$ the circular velocity depends on $r$. Integrating and
solving for the time yields
\begin{equation}
\label{eq_t-fric3}
t_\mrm{fric} = \frac{2 a_\mrm{h}}{12\pi S(\chi)\ln\Lambda}
\sqrt{\frac{N}{Gm_\ast}}\, \frac{\ln(x_\mrm{in}\eta)}{\ln\eta}\,x_\mrm{in}.
\end{equation}
Plotting $t_\mrm{fric}$ in the same way as $t_\mrm{diff}$ in
\fgr{f_t-diff} we obtain very similar curves whith only slightly
smaller times. Thus the relaxation time due to dynamical friction is
in very good agreement with the time obtained using diffusion
coefficients, confirming a larger relaxation time for steeper
profiles. For both results the relaxation time quickly drops to zero
as the inner boundary of the distribution is approached. This is just
a consequence of our assumption that the mass distribution does not
extend to smaller radii than $a_\mrm{g}$. However, real cusps will not
have such a sharp cut-off and their profiles will extend into the
inner regions, although with a slope less steep than $\gamma=3$ (the
mass has to be finite), resulting in times larger than those shown in
\fgr{f_t-diff}.
\begin{figure}
\begin{center}
\includegraphics[width=\FW]{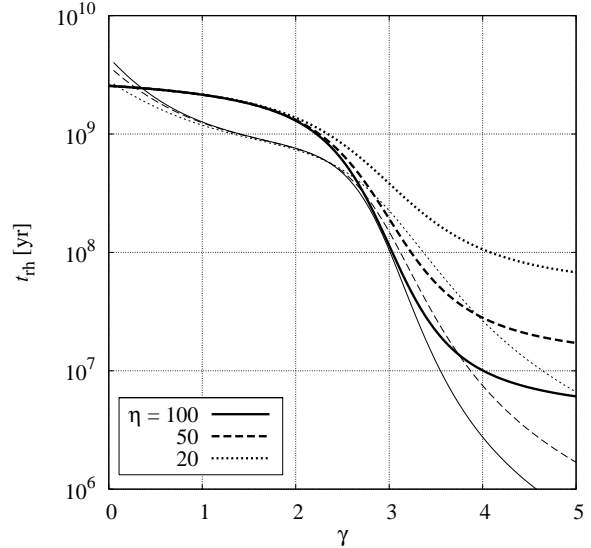}
\caption[]{The median relaxation time as a function of the cusp's
  slope. The bold curves have been derived using the diffusion
  coefficients while for the thin curves the dynamical friction
  formalism has been used, see \eqn{eq_th}.}
\label{f_th}
\end{center}
\end{figure}

To avoid these problems we try to characterize the cusp by a single
relaxation time, as suggested by \cite{binney94}, and replace the
density by the mean density $\rho_h =
\frac{1}{2}M_\mrm{c}/(\frac{4}{3} \pi r_h^3)$ inside the cusp's
half-mass radius $r_h$. The mean-square speed of the stars is best
described by $\langle v^2\rangle \approx 0.4 GM_\mrm{c}/r_h$, and we
use $\Lambda = r_h \langle v^2\rangle/Gm_\ast = 0.4 N$. For the
diffusion and friction time we then obtain
\begin{equation}
\begin{split}
t_{\mrm{diff},h} = & \frac{\sqrt{6}}{4\pi}\,\frac{\langle v^2\rangle^{3/2}}
  {G^2 m_\ast \rho_h \ln\Lambda}\\ t_{\mrm{fric},h} = & \frac{\langle
  v^2\rangle^{3/2}}{8\pi G^2 m_\ast S(\chi)\ln\Lambda}\,\frac{x_h^\gamma
  -\eta^{-\gamma}}{\gamma \rho_0},
\end{split}
\label{eq_th}
\end{equation}
where the half-mass radius in dependency of the slope is
\begin{equation}
x_h = \frac{r_h}{a_\mrm{h}} = \begin{cases}
  \left(\frac{\DS 1+\eta^{\gamma -3}}{\DS 2}\right)^{1/(3-\gamma)}, &
  \gamma \neq 3\\[2.5ex]  
  \frac{\DS 1}{\DS \sqrt{\eta}}, & \gamma = 3,
\end{cases}
\label{eq_xh}
\end{equation}
and $\rho_0$ is taken from \eqn{eq_rho0}. Note that for $\gamma =3$
the half-mass radius is the geometrical mean of $a_\mrm{g}$ and
$a_\mrm{h}$, i.e. $r_h = \sqrt{a_\mrm{g}a_\mrm{h}}$. The dependency of
these times on the slopes with $\eta$ as parameter is shown in
\fgr{f_th}.  On average the diffusion coefficients (bold lines) yield
slightly larger relaxation times than dynamical friction (thin
lines). Both times decrease with increasing slopes, as might be
expected, and increase with decreasing $\eta$, i.e. increasing
$a_\mrm{g}$. While for small $\gamma$ the relaxation time does not
sensitively depend on the slope it quickly decreases at $\gamma\approx
3$ before becoming flatter again at $\gamma\gtrsim 4$. For both
methods we find $t_\mrm{relax}$ to be in the range between $10^8$ and
$10^9\,\mrm{yr}$ for a slope in the range $2\lesssim\gamma\lesssim
3$. In the following we assume that a cusp with $\gamma\approx 3$ is
conserved for less than $10^9\,\mrm{yr}$.

The strict conservation of the slope $\gamma= 3$ is only necessary in
the theoretical derivation of \eqn{eq_rho}. We assumed the stars to
move on circular orbits and that the binary can only eject those stars
whose radius corresponds to the current semimajor axis. For elliptical
stellar orbits the distribution could be flatter and extend to larger
radii, but a more massive cusp would be required for coalescence due
to a smaller binding energy, as explained previously. Once the binary
has become hard it will influence all stars moving at smaller radii
than $a$. If the inner parts of the cluster collapse, maybe even into
a third \BH{}, its mass will become bound more deeply in the potential
of the binary and hence its ejection will extract more energy. Only if
these stars are so deeply bound in the potential of one of the \BH{}s
that the perturbations caused by the secondary are negligible the
merger might stop. Therefore the \BH{}s might still coalesce when the
time since they have become hard exceeds a couple of $10^8\,\mrm{yr}$
and a $\rho\propto r^{-3}$ profile is not conserved. We may conclude
that the binary covers the second phase when it decays from
$a_\mrm{h}$ to $a_\mrm{g}$ in the time $t_\mrm{hg} < 10^9\,\mrm{yr}$.



\section{Time dependency of the merger}
\label{s_shrinking-rate}
With our approach to compute the required stellar mass for a
successfully merging \BHB we can not determine the rate at which stars
are ejected and consequently no shrinking rate of the binary. To
derive the time dependency of the merger we would have to make further
assumptions about the rate at which stars interact with and are
ejected by the \BH{}s, i.e.  about the velocity distribution of the
stars. This might be quite difficult during an ongoing
merger. Nevertheless, one assumption we want to make is that the
shrinking rate is constant once the binary has become hard,
\begin{equation}
H\equiv \frac{\D}{\D t}\left(\frac{1}{a}\right) = \mathit{const.}
\label{eq_h-rate-def}
\end{equation}
This behaviour has been observed for hard binaries and a full loss-cone in
various numerical experiments \citep{hills83,quinlan96,milos01,zier01} and
should therefore also be applicable for steep and compact cusps and to
triaxial galaxies whose loss cone is always full
\citep{yu02,holley-bockelmann02,holley-bockelmann06,berczik06}. Note that our
definition is different from that used by \citet{quinlan96} who defined the
shrinking or hardening rate as a dimensionless quantity. We assume that the
binary needs the time $t_\mrm{hg}$ to shrink from the initial semimajor axis
$a_\mrm{h}$ to $a_\mrm{g}$. Integrating \eqn{eq_h-rate-def} in time from $t=0$
to some time $t<t_\mrm{hg}$ and in distance from $a_\mrm{h}$ at $t=0$ to
$a(t)$, respectively, we can solve for the semimajor axis which the binary has
reached after the time $t$ has elapsed and obtain
\begin{equation}
\frac{a(t)}{a_\mrm{h}} = \frac{1}{1+Ha_\mrm{h} t}.
\label{eq_aoft}
\end{equation}
If $t=t_\mrm{hg}$ the binary has shrunk to $a_\mrm{g}$ and solving for the
hardening rate yields $H = (\eta -1)/a_\mrm{h} t_\mrm{hg}$. Applying this to
\eqn{eq_aoft} we can rewrite the semimajor axis in the form
\begin{equation}
\frac{a(t)}{a_\mrm{h}} = \frac{1}{1 + (\eta -1) t/t_\mrm{hg}}.
\label{eq_a-t}
\end{equation}
The integration of \eqn{eq_da} gives the mass which has been ejected by the
time the semimajor axis has shrunk to $a(t)$ and we only need to replace
$a_\mrm{g}$ with $a(t)$ in \eqn{eq_mej} to obtain an expression for
$m(t)$. With the help of \eqn{eq_a-t} we can write the time dependency of the
ejected mass as
\begin{figure}
\begin{center}
\includegraphics[width=\FW]{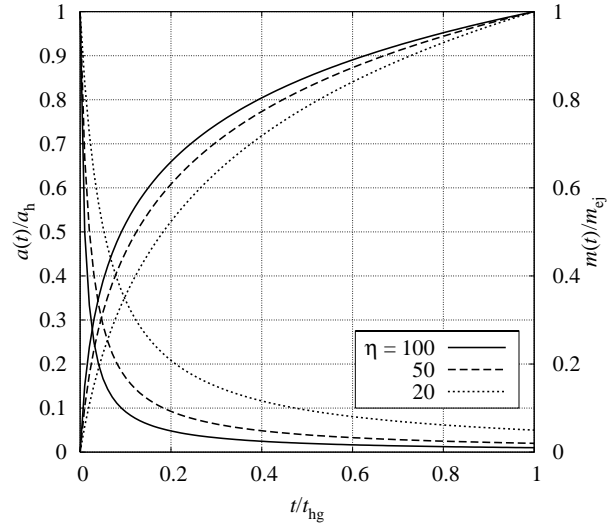}
\caption[]{Evolution of the semimajor axis (decreasing curves) and the ejected
mass (increasing curves) with time. The shrinking rate is assumed to be
constant.}
\label{f_a-m_t} 
\end{center}
\end{figure}
\begin{equation}
\frac{m(t)}{m_\mrm{ej}} = \frac{\ln\left[1 + (\eta -1)
    t/t_\mrm{hg}\right]}{\ln\eta}.
\label{eq_mej-t}
\end{equation}
Both functions are plotted in \fgr{f_a-m_t} for different ratios
$\eta$. The increasing curves show the mass evolution and the
decreasing ones show the change in the semimajor axis with time. In
the beginning the evolution of the binary is fastest, especially the
decay of the semimajor axis. It has decreased to $1/5$ or less of its
initial value after only $\sim 1/5$ of the merging time
$t_\mrm{hg}$. In the same period the ejected mass amounts to more than
a half of the total mass $m_\mrm{ej}$ which the \BH{}s have ejected
once they reach the separation $a_\mrm{g}$. After a time of about
$0.2\,t_\mrm{hg}$ has elapsed the evolution slows down noticably and
the binary spends most of its time in the second phase to shrink the
remaining distance from less than $0.2\, a_\mrm{h}$ to $a_\mrm{g} =
a_\mrm{h}/\eta$. Hence, if a binary really stalls it will be most
likely found in the range $a_\mrm{g}\lesssim a\lesssim
0.2\,a_\mrm{h}$, see Section \ref{s_ongoing}. This is in agreement
with the results of the numerical three-body experiments by
\citet{roos81} which suggest that the binary stops shrinking at a
separation of the \BH{}s of about $0.015\,r_\mrm{cusp}$.

Initially the mass ejection rate is $\dot{m}(0)=\mu (\eta -1)/(k
t_\mrm{hg})$ and exceeds the rate at the end of the second phase when
$t=t_\mrm{hg}$ by a factor of $\eta$. This is probably the time when
the \BHB carves a torus comprised of stars out of the initial stellar
distribution \citep{zier01,zier02}. In Section~\ref{s_trelax} we
showed that the steep profile of the cusp is conserved for some
$10^8\,\mrm{yr}$, and therefore the binary should shrink from
$a_\mrm{h}$ to $a_\mrm{g}$ on a similar time scale or less. Taking
$t_\mrm{hg} = 5\times 10^8\,\mrm{yr}$ and assuming a mass ratio $q=1$
with $m_1\approx 10^8\,M_\odot$ we obtain an initial ejection rate of
about $2$, $5$ and $10\,M_\odot/\mrm{yr}$ for $\eta= 20$, $50$ and
$100$, respectively. At the end of phase 2 the ejection rate is smaller
by a factor of $1/\eta$ and amounts only to $0.1\,M_\odot/\mrm{yr}$,
being almost independet of $\eta$ for $\eta\gg 1$. This again
indicates that the evolution at the end of phase 2 is much slower than
in the beginning.

We have not taken into account any dependency on the density profile
and velocity distribution. While the shrinking rate might still be a
constant due to a filled loss-cone, the shrinking time $t_\mrm{hg}$
probably depends on both distributions. This could be explored in more
detail with the help of $N$-body simulations. From our estimates in
Section~\ref{s_trelax} of the relaxation time of the cusp we deduced a
time $t_\mrm{hg}$ less than $10^9\,\mrm{yr}$. Another estimate for the
order of magnitude of $t_\mrm{hg}$ could also be derived from
\Z-shaped radio galaxies, as has been pointed out by \citet{zier05},
but that is beyond the scope of the present paper.


\section{\lowercase{$m_\mrm{ej}$} in multiple mergers}
\label{s_multiple}
Although our approach to the merger of a massive \BHB is quite simple
the results seem to describe such a merger reasonably well. However,
there is a simple and easy consistency check with numerical
simulations which we want to carry out.  According to simulations more
mass is ejected if the primary BH merges $N$ times with a BH of mass
$m_1/N$ than in one merger with a secondary BH of mass $m_2=m_1$
\citep[e.g.][]{quinlan96,zier01}. In case of one merger we have
$m_\mrm{ej} = m_1\ln(a_\mrm{h}/a_\mrm{g})/(2k) $, cf. \eqn{eq_mej}
with $q=1$. If we distribute the merging mass over $N$ mergers we have
$m_2 = m_1/N$. When the $i$th merger proceeds, the primary's mass is
that it has after the $(i-1)$th merger has been completed,
\begin{equation}
m_{1,i} = m_1 + \frac{i-1}{N} m_1 = m_1 \frac{N+i-1}{N}.
\label{eq_m1i}
\end{equation}
The mass ratio during the $i$th merger then is
\begin{equation}
q_i = \frac{m_2}{m_{1, i}} = \frac{1}{N+i-1},
\label{eq_qi}
\end{equation}
and hence the reduced mass $\mu_i = m_{1,i}\,q_i/(1+q_i)$.  Thus we can write
the mass which becomes ejected during the $i$th merger as
\begin{equation}
m_{\mrm{ej},i} = \frac{m_1}{k} \ln\left(\frac{a_\mrm{h}}{a_\mrm{g}}\right)
\frac{1}{N}\frac{N+i-1}{N+i}.
\end{equation}
The total mass ejected in $N$ mergers then amounts to
\begin{equation}
m_\mrm{ej,tot} = \sum_{i=1}^{N} m_{\mrm{ej},i} = \frac{m_1}{k} \ln
\left(\frac{a_\mrm{h}}{a_\mrm{g}}\right) \frac{1}{N}\sum_{i=1}^{N}
\frac{N+i-1}{N+i}.
\end{equation}
The sum on the right hand side is equal to $N + \psi(1+N) - \psi(1+2N)$, where
$\psi$ is the Psi (Digamma) Function and can be approximated by $N^2/(1+N)$,
with a maximum error of about $6\%$ for $N= 2$. This approximation corresponds
to keeping $i$ fixed to $1$ in the above expression for $m_{1,i}$ so that
$m_{1,i}$ remains constant at $m_{1,i}=m_1$ for all mergers. Hence the
primary's mass is fixed and its growth with the increasing number of mergers
can be neglected. The total ejected mass after $N$ mergers is approximately
\begin{equation}
m_\mrm{ej,tot} \approx \frac{m_1}{k} \, \frac{N}{1+N} \ln
\frac{a_\mrm{h}}{a_\mrm{g}}
\end{equation}
and the ratio of the mass ejected in $N$ mergers with $m_2 = m_1/N$ compared
to one merger with $m_2 = m_1$ is
\begin{equation}
\frac{m_\mrm{ej,tot}}{m_\mrm{ej}} = 2\,\frac{N}{1+N}.
\label{eq_mratioN}
\end{equation}
This is a function that grows with $N$, in agreement with the results
of numerical simulations. In deriving this ratio we neglected the
dependency of $\eta = a_\mrm{h}/a_\mrm{g}$ on the mass ratio
$q$. According to numerical experiments this ratio is increasing with
decreasing $q$. Therefore, the inclusion of this dependency would
result in an ejected mass which increases more steeply with $N$,
making our result more pronounced.


\section{Ongoing mergers}
\label{s_ongoing}
The results we obtained in the present article and \citetalias{zier06}
suggest that the \BHB which forms after the collision of two galaxies
most likely merges. In the introduction we cited observational
evidence for ongoing mergers where the \BH{}s are still orbiting
around each other. It has been pointed out by \citet{gopal-krishna03}
that in ZRGs during the time between the bending of the pre-merger jet
into a \Z-shape by the secondary galaxy and the launching of the
post-merger jet after the coalescence of both \BH{}s we should see
only the pure \Z-shape, but no complete \X-shape of the jets. So far
no galaxies with a pure \Z-shape have been observed. In the new and
strongly increased sample of XRGs, compiled by \citet{cheung07}, about
three such sources out of hundred galaxies might have been observed
for the first time, i.e. J0145-0159, J1040+5056 and J1206+3812. Of
course this needs a thorough and detailed analysis. However, even the
existence of these objects would not indicate that the binary has
stalled and is rather a sign that the separation of the \BH{}s is
somewhere below $30$-$100\,\mrm{kpc}$ \citep{zier05}, therefore
providing a very important laboratory for the research of ongoing
mergers and merger history. The sample of \citeauthor{cheung07} also
seems to increase the number of post-merger ZRGs, supplying more
objects for the deprojection of the jets as has been done by
\citet{zier05}. The pure \Z-shaped sources might be good candidates to
look for a spatially resolved binary like in NGC~6240 which has been
discovered by \citet{komossa03a}. Because of the projected separation
of the \BH{}s of about $1.4\,\mrm{kpc}$ this merger is currently in
the first phase. More recently \citet{rodriguez06} have discovered a
black hole binary with a total mass of $\sim 1.5\,10^8\,M_\odot$ and a
projected separation of only $7.3\,\mrm{pc}$ in the radio galaxy
0402+379, probably the most tightly bound binary that has been
observed directly so far. The elliptical host galaxy shows signs of a
recent merger and the two nuclei appear to be active, suggesting
ongoing accretion and dissipation. Because the separation is seen in
projection it is rather a lower limit and therefore the merger is
still in the first phase or at most at the beginning of the second
when the loss-cone is still full. Hence, this is another example of an
ongoing merger where the loss-cone is not depleted. However, based on
numerical simulations of a \BHB in a stellar core some authors claimed
that probably the binary stalls in the second phase due to loss-cone
depletion
\citep[e.g.][]{quinlan96,makino97,quinlan97,milos01,berczik05}. To
test this prediction, which contradicts ours of successfully
coalescing \BH{}s, we look for ongoing mergers which happen to be in
the second phase. These might still not be stalled and actually en
route to coalescence. Only if there is a significant number of ongoing
mergers in the second phase, preferentially in the range
$a_\mrm{g}\lesssim a\lesssim 0.2\,a_\mrm{h}$ as we predicted in
Section~\ref{s_shrinking-rate}, this could indeed argue for a stalled
binary. A still existing \BHB manifests itself also in semi-periodic
signals in lightcurves, see \citet{komossa03b,komossa06} and
references therein.  \citet{katz97} presented a model for OJ~287 where
the precession of the accretion disc, driven by the gravitational
torque of the secondary \BH, causes the jet to sweep periodically
across our line of sight. Doppler-boosting leads to the observed
variations of the luminosity and the period of the binary is much less
than the $9\,\mrm{yr}$ interval of the luminosity variations in the
rest frame of the galaxy. Models by \citet{sillanpaa88} and
\citet{valtaoja00} relate the variations of the lightcurve to
interactions of the secondary \BH with the accretion disc and
therefore the observed period corresponds to the orbital period of the
binary. \citet{merritt05} have compiled a sample of active galaxies
whose observed periodic variabilities might be related to the orbital
motion of the \BH{}s. We assume that the intervals of these
variabilities correspond to the period of the binary and use them
together with independent estimates for the mass of the \BH{}s to
determine the separation of the \BH{}s and hence the phase in which
the merger has been observed. Kepler's third law relates the binary's
period $T$ to its semimajor axis,
\begin{table*}
\caption[]{Sources exhibiting (semi)periodic changes in lightcurves,
  possibly due to a \BHB. Columns: (1) source, (2) redshift, (3)
  intrinsic period, (4)-(7) \BH masses, (8) current separation scaled
  to $a_\mrm{g}$ in units of $10^{-3}$ ($q=0.1$), (9) mass ratio
  obtained under the condition that $a = a_\mrm{g}$, and (10)
  remaining time to coalesce due to emission of gravitational waves
  ($q=0.1$).}
\label{t_one}
\begin{tabular*}{\textwidth}[]{@{\extracolsep\fill}lrrllllccc@{}}
\hline\\[-1ex]
Source & \multicolumn{1}{c}{$z$} & \multicolumn{1}{c}{$T_\mrm{intr}$} &
\multicolumn{4}{c}{$\log(M_\mrm{BH}/M_\odot)$} &
\multicolumn{1}{c}{$a/a_\mrm{g}$} &
\multicolumn{2}{c}{$a=a_\mrm{g}$}\\[0.5ex]

& & \multicolumn{1}{c}{[yr]} & & & & &
 \multicolumn{1}{c}{[$10^{-3}$]} & \multicolumn{1}{c}{$-\log(q)$} &
 \multicolumn{1}{c}{$\log (t_\mrm{g}/\mrm{yr})$} \\[0.5ex]

\multicolumn{1}{c}{(1)} & \multicolumn{1}{c}{(2)} & \multicolumn{1}{c}{(3)} &
 \multicolumn{1}{c}{(4)} & \multicolumn{1}{c}{(5)} & \multicolumn{1}{c}{(6)} &
 \multicolumn{1}{c}{(7)} & \multicolumn{1}{c}{(8)} & \multicolumn{1}{c}{(9)} &
 \multicolumn{1}{c}{(10)} \\[.75ex]

\hline\\[-2ex]

Mrk~421 & $0.031$ & $22.4$ & & & $7.6$ & $8.3^{\sigma}$ & $373\,\text{-}\,730$
& $2.7\,\text{-}\,1.6$ & $8.3\,\text{-}\,9.5$\\

Pks~0735+178 & $0.424$ & $10.0$ & $8.1$ & $8.3$ & $8.2$ & &
$218\,\text{-}\,264$ & $3.7\,\text{-}\,3.3$ & $7.4\,\text{-}\,7.7$\\

BL~Lac & $0.069$ & $13.1$ & $6.4$ & $7.3$ & $7.7$ & $8.4^{\sigma}$ &
$237\,\text{-}\,1615$ & $3.5\,\text{-}\,0.01$ & $7.5\,\text{-}\,10.8$\\

On~231 & $0.102$ & $12.3$ & & & $8.0$ & & $334$ & $2.9$ & $8.1$\\

Oj~287 & $0.306$ & $9.1$ & $7.7$ & $8.4$ & $8.1$ & $8.8^{\lambda}$ &
$127\,\text{-}\,364$ & $4.6\,\text{-}\,2.8$ & $6.4\,\text{-}\,8.3$\\

Pks~1510-089 & $0.361$ & $0.7$ & & & $8.0$ & $8.6^{\lambda}$ &
$27.8\,\text{-}\,49.4$ & $7.2\,\text{-}\,6.2$ & $3.8\,\text{-}\,4.8$\\

3C~345 & $0.595$ & $6.3$ & & & $8.0$ & $9.3^{\lambda}$ & $61.4\,\text{-}\,214$
& $5.9\,\text{-}\,3.7$ & $5.2\,\text{-}\,7.3$\\

AO~0235+16 & $0.940$ & $2.9$ & $8.7$ & $8.7$ & $8.0$ & & $65.1\,\text{-}\,127$
& $5.8\,\text{-}\,4.6$ & $5.3\,\text{-}\,6.4$\\

3C~66A & $0.444$ & $0.125$ & & & $8.0$ & & $15.7$ & $8.2$ & $2.8$\\

Mrk~501 & $0.033$ & $0.063$ & & & $8.3$ & $9.2^{\sigma}$ &
$3.14\,\text{-}\,7.43$ & $11\,\text{-}\,9.5$ & $-0.01\,\text{to}\,1.5$\\

3C~273 & $0.158$ & $0.00225$ & & & $9.0$ & $9.2^{L}$ & $0.34\,\text{-}\,
0.41$ & $15$ & $-3.9\,\text{to}\,-3.5$\\

Sgr A$^\ast$ & 0.0 & 0.3 & & $6.5^{(e)}$ & $6.6^{(f)}$ & &
$108\,\text{-}118\,$ & $4.9\,\text{-}\,4.7$ & $6.1\,\text{-}\,6.3$\\[.75ex]

\hline
\end{tabular*}

\medskip
\begin{minipage}{\textwidth}
\emph{References for the periods:} Mrk~421 \citep{liu97}, Pks~0735+178
\citep{fan97}, BL~Lac \citep{fan98}, On~231 \citep{liu95}, Oj~287
\citep{pursimo00}, Pks~1510-089 \citep{xie02b}, 3C~345 \citep{zhang98},
AO~0235+16 \citep{raiteri01}, 3C~66A \citep{lainela99}, Mrk~501
\citep{hayashida98}, 3C~273 \citep{xie99}, Sgr~A$^\ast$
\citep{zhao01}. \emph{References for the masses:} With the exception of
Sgr~A$^\ast$ the values in columns (4) and (5) were taken from \citet{xie02},
in column (6) from \citet{xie04} and in column (7) from \citet{wang04}. Masses
for Sgr~A$^\ast$ are from $\phantom{}^{(e)}$ \citet{schoedel03} and
$\phantom{}^{(f)}$ \citet{ghez03}. The indices $\sigma,\,\lambda,\,L$ indicate
the method used to determine the \BH mass. See text for details.
\end{minipage}
\end{table*}
\begin{equation}
T = 2\pi \sqrt{\frac{a^3}{GM_{12}}}.
\label{eq_period}
\end{equation}
We scale $a$ with the semimajor axis where gravitational waves start
to dominate the decay of the binary, $a_\mrm{g}$. For circular orbits
this is \citep{peters64}
\begin{equation}
a_\mrm{g} = \left[\frac{256}{5}\,\frac{G^3\mu M_{12}^2}{c^5}\,
  t_\mrm{g}\right]^{1/4}.
\label{eq_ag}
\end{equation}
Note that \citet{peters64} actually gave this expression in the form
$\langle \D a/\D t\rangle = -(64/5)\,G^3\mu M_{12}^2/(c^5 a^3)$ which
integrated yields \eqn{eq_ag}. Some authors alternatively used the
definition $t_\mrm{g}\equiv |\dot{a}/a|^{-1}$ evaluated at
$a=a_\mrm{g}$ which differs form $t_\mrm{g}$ in \eqn{eq_ag} by a
factor of $4$. Combining Eqs.~(\ref{eq_period}) and (\ref{eq_ag}) we
obtain for the current semimajor axis in units of $a_\mrm{g}$:
\begin{equation}
\frac{a}{a_\mrm{g}} = \frac{1}{29} \left( \frac{m_1}{10^8\,M_\odot}
\right)^{-\frac{5}{12}} \left( \frac{t_\mrm{g}}{10^{10}\,\mrm{yr}}
\right)^{-\frac{1}{4}} \left( \frac{T}{\mrm{yr}} \right)^{\frac{2}{3}}
\frac{(1+q)^{\frac{1}{12}}}{q^{1/4}}.
\label{eq_a}
\end{equation}
Because of the small exponent $1/4$ the semimajor axis depends only
weakly on the time. Scaling $t_\mrm{g}$ with $10^8\,\mrm{yr}$ instead
of $10^{10}\,\mrm{yr}$ increases $a$ by only a factor of $\sim 3$.
The period $T$ and the mass $m_1$ of the primary \BH of the binary we
derive from observations. Assuming a mass ratio $q$ then allows us to
compute the separation of the \BH{}s. In Table~\ref{t_one}, we listed
the sources in column (1) and used their redshifts (2) to transform
the observed periods into the rest frame of the source via
$T_\mrm{intr} = T_\mrm{obs}/(1+z)$, column (3). For the central \BH
mass we found different values in the literature which have been
obtained with various methods (columns (4) to (7)). \citet{xie02}
assume a maximally rotating Kerr \BH and relating the observed minimal
timescales of the luminosity variations, on scales between $1/2$ to
$12$ hours, to the period of the marginally bound orbit they obtain an
upper limit of the \BH mass which is given in column (5). The same
authors used an expression for the Eddington-limit that includes the
Klein-Nishina effects on the Compton scattering cross Section to
obtain a lower limit for the mass, listed in column (4), a method
proposed by \citet{dermer95}. Later \citet{xie04} used again the
method of the minimal timescales for a larger sample with the results
shown in column (6). \citet{wang04} employed three different methods
to determine the central \BH mass which is indicated by the upper
index in column (7). They either used the $M_\mrm{BH}-\sigma$ relation
as fitted by \citet{tremaine02} ($\sigma$), the
linewidth-luminosity-mass scaling relation \citep{kaspi00}
($\lambda$), or the correlation between the luminosity of the host
galaxy and \BH mass \citep{mclure01} ($L$). For Sgr~A$^\ast$
\citet{ghez03} and \citet{schoedel03} used observations of absorption
lines to determine the orbits of central stars and hence the mass of
the \BH.

As can be seen in Table~\ref{t_one} the masses for some objects are
quite different by up to a factor of $100$ (BL~Lac, 3C~345,
Oj~287). There is also a contradiction for AO~0235+16 between the
lower and upper limits of the mass in columns (4) and (6), respectively,
and between coulmns (6) and (7), where the latter mass is not supposed
to be an upper limit. Notwithstanding these problems we just took the
smallest and largest mass and computed the corresponding range of the
current separation of the \BH{}s using \eqn{eq_a} with $q=0.1$. The
result is given in column (8) in units of $10^{-3}$. For $q=1$ the
obtained semimajor axes would be even smaller by a factor of
$0.6$. Assuming that the binary's current semimajor axis is just at
the transition from phase 2 to 3, i.e. $a=a_\mrm{g}$ we can solve
\eqn{eq_a} for the required mass ratio $q$ and obtain
\begin{equation}
q = \frac{3^{1/3} 2 \chi + 2^{1/3} (\sqrt{3\chi^2 (27-4\chi)} +
    9\chi)^{2/3}}{6^{2/3}(\sqrt{3\chi^2 (27-4\chi)}+9\chi)^{1/3}}.
\end{equation}
Here we used the definition
\begin{equation}
\chi^{1/12} \equiv (4\pi^2)^{-1/3} c^{5/4} (G m_1)^{-5/12} T^{2/3}
t_\mrm{g}^{-1/4}.
\end{equation}
The negative logarithm of the mass ratio is tabulated in column
(9). Binaries with a smaller mass ratio are still in phase 2, while
for larger ratios gravitational radiation already dominates the
decay. In the last column we listed the remaining time for the \BH{}s
to merge due to emission of gravitational waves if $q=0.1$, which is
obtained by solving \eqn{eq_a} for $t_\mrm{g}$ with $a=a_\mrm{g}$,
\begin{equation}
t_\mrm{g} = 1.5\,\,10^4 \left(\frac{m_1}{10^8\,M_\odot}\right)^{-5/3}
\left(\frac{T}{\mrm{yr}}\right)^{8/3} \frac{(1+q)^{1/3}}{q}\,\mrm{yr}.
\label{eq_t}
\end{equation}
For $q=1$ this time is shorter by a factor of about $0.12$.
\begin{figure}
\begin{center}
\includegraphics[width=\FW]{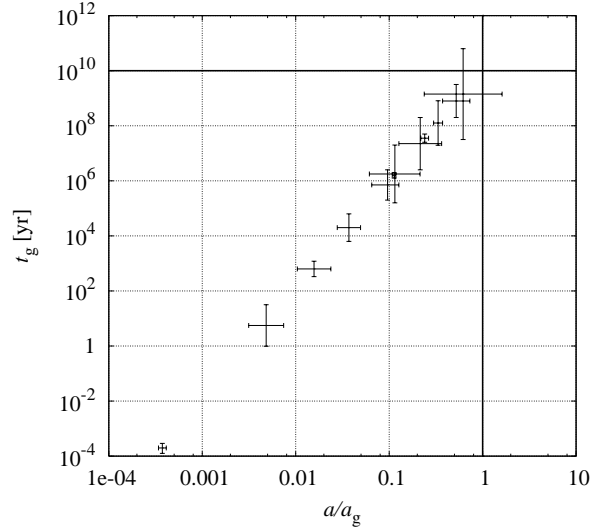}
\caption[]{Plot of column (8) vs. (10) of Table~\ref{t_one} for
  $q=0.1$. Data lie on the curve defined by \eqn{eq_a} with the
  errorbars corresponding to the mass ranges in column (4) to
  (7). Horizontal and vertical lines indicate transition
  between phases 2 and 3.}
\label{f_t-data}
\end{center}
\end{figure}

The results show very clearly that all binaries without exception are
already in the third phase of the merging process. This can also be
seen in \fgr{f_t-data} where we plotted column (8) vs. (10). Only if
we use the lower limit for the mass of BL~Lac the separation of the
\BH{}s is larger by a factor of $1.6$ than $a_\mrm{g}$ and
consequently the remaining merging time due to emission of
gravitational waves exceeds a Hubble time (columns (8) and
(10)). Taking the mass from column (7), which is obtained from the
$M_\mrm{BH}\,\text{--}\,\sigma$ relation and therefore might be more
reliable, also this source is well beyond the limit to the third
phase. For the intermediate masses, $\log(M_\mrm{BH}/M_\odot) = 7.7$
and $7.3$, we obtain $a/a_\mrm{g} = 0.46$ and $0.68$, respectively. As
said before a larger mass ratio would further diminish the current
separation of the \BH{}s. Thus, even for the lower limit of the mass
of BL~Lac an equal mass binary would be in the third phase. The
smaller the mass ratio is, the larger is the current separation and
the longer it takes for the \BH{}s to merge. From column (9), showing
the negative logarithm of the mass ratio for which the separation is
$a=a_\mrm{g}$ and the remaining merging time is the Hubble time, we
see that even for very small mass ratios the binary just enters the
third phase. All upper limits are smaller than $0.1$. Of course BL~Lac
again is the sole exception, but only for the lower mass limit. All
binaries, with the exception of BL~Lac for the small mass limit, will
coalesce in much less than a Hubble time (column (10)). Using $q=1$
instead of $0.1$ the merging times are smaller by a factor of $0.12$
so that the binary in BL~Lac will merge in less than a Hubble time
also for the small mass limit. In case of eccentric orbits of the
\BH{}s the merging times would be further decreased. Thus the times we
obtained are actually upper limits. Explaining the variations in the
lightcurves with the model by \citet{katz97} would result in much
smaller periods of the binaries and hence in smaller separations and
merging times (Eqs.~(\ref{eq_a}, \ref{eq_t})). Consequently the binary
would be even deeper in the third and last phase of the merger.
Therefore, provided that the variations in the lightcurves are due to
a \BHB in the center of the galaxies, our findings in this Section are
in very good agreement with the results of the previous Sections and
strongly support our conclusions in \citetalias{zier06}: Most likely
the slingshot mechanism in the second phase of the merger is
sufficiently efficient in order to extract enough energy and angular
momentum from the binary so that the \BH{}s can enter the final
phase. Hence the profile of the cusp at the beginning of the second
phase is very steep indeed, and the binary covers this phase in less
than $\sim 10^9\,\mrm{yr}$ (Section ~\ref{s_mass-dist}). Once
gravitational radiation dominates the shrinking in the final phase the
\BH{}s merge in less than a Hubble time. It is actually striking that
all possible non-merged \BHB{}s are observed in all phases but the
second, which according to loss-cone depletion models is the one in
which binaries should most likely be found.


\section{Discussion and Conclusions}
\label{s_summary}
Observational evidence suggests that \BHB{}s, formed in a galaxy
collision, eventually coalesce within less than a Hubble
time. Focusing on stars bound to the binary we showed in
\citetalias{zier06} that slingshot ejection of stars in the second
phase of a merger, which is considered to be the bottleneck, is
sufficiently efficient to allow the \BH{}s to coalesce. The
prerequisite is a steep cusp which is about as massive as the binary
when the binary becomes hard. In this paper, we further pursued this
idea and compared its predictions with observations and numerical
simulations. Our results verify and strengthen our conclusion of a
steep cusp and that the \BH{}s coalesce in less than a Hubble time.

In Sections~\ref{s_k} and \ref{s_potential}, we examined in detail the
assumptions on which our results in \citetalias{zier06} are based: The
kick-parameter is about $1$ and we can neglect the cluster potential
when calculating the energy which the binary loses to the ejected
stars. Our crude theoretical estimate for $k$ is in agreement with the
kick-parameter we derived from the data obtained from simulating a
stellar cluster in the potential of a binary moving on fixed orbits
\citep{zier00}. Comparing these values with those obtained in other
simulations and scattering experiments
\citep{hills80,roos81,quinlan96,yu02} we found very good agreement,
justifying our assumption of $k\approx 1$. While including the cluster
potential tends to slightly increase the kick-parameter on average, we
found that it can be neglected in comparision with the potential of
the binary when computing the energy of the stars. The influence of
the cusp on the total potential becomes even less for steeper profiles
which are required for a successful merger. Hence the assumptions are
well justified and we can be confident in our results.

In Section~\ref{s_density}, we derived a density profile which enables
the \BH{}s to merge and is in agreement with the observed post-merger
profiles.  Using a deprojected S\'{e}rsic model without a cusp for
non-core galaxies results in a large amount of total mass for a
successful merger if we use a shape parameter $n\approx 5$, which has
been found by \citet{trujillo04} for the best fits (cf. model~(a) in
\fgr{f_m-gal}). If, however, a steep central cusp is formed which is
removed subsequently by the binary in the second phase, like in
model~(b), the total mass is much less. To obtain the initial profile,
we add such a cusp to a post-merger profile of a core galaxy, best
approximated by a core-S\'{e}rsic model with $n\approx 5$
(\citeauthor{trujillo04}). For $\gamma\gtrsim 2.5$ this pre-merger
profile allows the \BH{}s to merge without the total mass becoming too
large ($\sim 10^{12}\,M_\odot$), with a core galaxy as merger remnant,
see \fgr{f_m-cgal}. Without this cusp, i.e. for a flat core with
$\gamma \lesssim 2$, either a very large amount of mass is needed, or
the binary stalls before it is able to enter the third phase. This is
in good agreement with the results from numerical three-body
experiments obtained by \citet{roos81}. Hence we argue that a large
and shallow stellar cluster is the end product of a merger while at
the time the binary becomes hard a steep cusp is formed which allows
the \BH{}s to coalesce. The mass distribution of a core galaxy with a
cusp is shown in \fgr{f_rho-nc} by the thin dash-dotted line. It
follows a flat power law between $a_\mrm{h}$ and the break radius
$r_b\approx 10\,a_\mrm{h}$. If core galaxies are merger remnants and
the core is formed by mass ejection of the binary we would expect
$a_\mrm{h}$ to coincide with $r_b$. This discrepancy might be caused
by the assumption of a too small $a_\mrm{h}$: Stars bound very tightly
to their host \BH might increase its effective mass and hence the
second phase might start earlier at larger distances.  Furthermore we
only considered circular stellar orbits. Stars moving on eccentric
orbits with apocenters in the range $a_\mrm{h}\lesssim r_+ \lesssim
r_b$ and pericenters $r_-\lesssim a_\mrm{h}$ also interact with the
binary. The ejection of this additional mass will shift $a_\mrm{h}$ to
larger radii, up to $r_b$ for $\epsilon\approx (r_b - a_\mrm{h})/(r_b
+ a_\mrm{h})$. Such stars do not extract as much energy as the more
tightly bound stars moving on circular orbits and more mass has to be
ejected. However, these stars could compensate for this mass and
enable successful mergers in shallower cusps.

Depending on how closely the profile of the ejected mass approaches
the initial distribution all slopes for the final profile which are
less than that of the cusp are possible. This even includes
distributions where the density drops with decreasing radius, which
have actually been observed by \citet{lauer02}. They might indicate
that the binary got stalled \citep{zier01}, but can also be formed by
\BH{}s which successfully merge. The maximum of these distributions is
observed to be at radii about a factor of $10$ larger than
$a_\mrm{h}$, i.e. close to $r_b$ of core galaxies, and thus supports
the arguments above for a larger $a_\mrm{h}$.

At the end of Section~\ref{s_density} we computed the ralaxation time
of the cusp in dependency of its slope and obtained a couple of
$10^8\,\mrm{yr}$ if $\gamma = 3$. For longer times the steep profile
is not conserved and if the binary could only eject those stars whose
orbital radius is the same as the current semimajor axis, as we
assumed to derive \eqn{eq_rho}, the merger would stall. Of course the
\BH{}s, once they have become hard, will influence all stars at
$r\lesssim a$, and only if they are very deeply bound in the potential
of one of the \BH{}s they will not be ejected. Hence deviations from
$\rho\propto r^{-3}$ will not immediately cause the binary to
stall. Deriving \eqn{eq_rho} we assumed the stars to move on circular
orbits. As we pointed out above the inclusion of eccentric orbits
allows for shallower ejected profiles ($\gamma\lesssim 3$) which might
extend to $r_b$ instead of $a_\mrm{h}$. The shallower cusps would have
a larger relaxation time, see \fgr{f_th}. Thus we expect its lifetime
between $10^8$ and $10^9\,\mrm{yr}$ for $3\lesssim\gamma\lesssim 2.5$
and the \BH{}s to decay from $a_\mrm{h}$ to $a_\mrm{g}$ on the time
scale $t_\mrm{hg}$ which is similar or less.

All observational evidence for successfully merged \BH{}s, cited in
the introduction, also supports our prediction of the transient
formation of a steep cusp. But how is such a cusp formed? Adding
adiabatically growing \BH{}s to nonrotating spherical galaxy models
and seeking for equilibrium solutions for the cusp, \citet{quinlan95}
find slopes as steep as $2.5$ with an initial $\gamma = 2$ and argue
that steeper cusps without central \BH{}s are unlikely. However, the
steep cusp which is required for the merger is a transient feature and
does not need to be in a state of equilibrium. Indeed, Fig.~3 in
\citet{milos01} suggests that immediately after the merger of both
cores when the binary becomes hard the density profile inside
$a_\mrm{h}$ is substantially steeper than $\gamma = 2$. Whether this
is a real physical effect or due to spurious numerical relaxation
still needs to be clarified. When both galaxies merge energy will be
dissipated and fractions of angular momentum cancel each other. The
amount of these fractions will depend on the initial mass and velocity
distributions in the isolated galaxies and on the magnitude and
orientation of both galactic spins and the orbital angular momentum
relative to each other. The mass that is funneld into the common
center of the galaxies will merge with the cores surrounding each
\BH. At this time the central potential is strongly nonspherical so
that both cores will be heavily perturbed, being far from a state of
equilibrium, with profiles that might be steeper than $\gamma =2$. At
$a=a_\mrm{h}$ we expect them to merge into the required steep cusp
with a mass of $\sim M_{12}$. At this stage of the merger we can only
speculate about the central mass distribution which might be traced
with the help of detailed numerical simulations taking into account
the above mentioned initial conditions. They will also influence the
morphology of the merger product \citep{toomre72} as well as processes
like the star formation rate and hence the final gas and star
content. After the coalescence of the \BH{}s the velocity of the
surrounding stars will be tangentially anisotropic at $r\lesssim
a_\mrm{h}$ and radially anisotropic at larger radii, with the ejected
stars being focused to the equatorial plane of the binary, the more
the larger their kinetic energy is \citep{zier01}. However, if the
central regions of the cusp undergo a core collapse before the \BH{}s
could merge, possibly leading to the formation of a third \BH{}, this
mass is still available for slingshot ejection by the \BH{}s which
therefore still can enter the final phase. Our derivation of the
initial profile might be oversimplified. A fraction of the ejected
stars will stay bound to the cluster at larger distances while the
stars remaining in the cusp region are less tightly bound due to the
smaller mass in the center and hence will also expand to larger
regions. This is enhanced by the energy transfer from the binary to
these stars, which are still bound by the \BH{}s, i.e. the heating of
this population. Hence, there is a shift of mass within the cluster
from the inner to the outer regions, resulting in a profile which is
flatter than the difference between the initial and ejected mass
distribution. However, the mass deficiency which has been derived by
\citet{graham04} from the difference between the core-S\'{e}rsic fit
and the extrapolation of the pure S\'{e}rsic profile into the central
region amounts to $1\text{--}2\,M_{12}$. This is in very good
agreement with our results, provided that $\gamma\gtrsim 2.5$, see
\fgr{f_m-def}.

Although our approach to the merging of the \BH{}s does not allow to
determine the time dependency, we made use of the finding in numerical
experiments that the hardening rate is constant once the binary has
become hard \citep{hills83,quinlan96,milos01,zier01}, see
Section~\ref{s_shrinking-rate}.  Utilizing this assumption we obtained
the semimajor axis and the ejected mass as functions of time with the
parameter $t_\mrm{hg}$, the time to dacay from $a_\mrm{h}$ to
$a_\mrm{g}$. This is less than $10^9\,\mrm{yr}$ as we concluded from
the relaxation time of the cusp in Section~\ref{s_trelax}. The exact
value of this shrinking time will depend on the mass and velocity
distribution of the stars and hence on the initial conditions of the
merger. Our results showed that the binary elvolves fastest in the
beginning of phase 2 and then continuously slows down. Therefore, we
conclude that if stalled binaries exist at all, they will most likely
be found with a semimajor axis in the range $a_\mrm{g} \lesssim a
\lesssim 0.2\,a_\mrm{h}$. This is in agreement with the results of
numerical three-body experiments by \citet{roos81} who finds that a
binary might stall at $a\approx 0.015\,r_\mrm{cusp}$.

In Section~\ref{s_multiple}, we performed a simple consistency
check. In agreement with numerical simulations
\citep{quinlan96,quinlan97,milos01,zier01} we find that the ejected
mass is increasing with the number $N$ of mergers if we keep the total
mass which merges with $m_1$ constant. This means that more mass is
ejected in $N$ mergers of $m_1$ with $m_2 = m_1/N$ than in one merger
with $m_2 =m_1$. The stronger dependency on $N$ which has been found
in the simulations might be due to our assumption that we can neglect
the dependency of the ratio $a_\mrm{h} / a_\mrm{g}$ on the mass ratio
$q$. Our results show that the growth of $m_1$ during the $N$ mergers
has a negligible influence on the total ejected mass.

Ongoing mergers have been observed directly by \citet{komossa03a} in
NGC~6240 with a projected separation of $1.4\,\mrm{kpc}$, and by
\citet{rodriguez06} in the elliptical galaxy 0402+379 with a projected
separation of both nuclei of only $7.3\,\mrm{pc}$. Thus both mergers
are still in the first phase and are clearly far from being stalled at
the end of the second phase in a distance of a few
$0.01\,\mrm{pc}$. Other promising sources with still existing binaries
could be pure \Z-shaped radio galaxies \citep{gopal-krishna03} with a
separation less than $30 \text{-} 100\,\mrm{kpc}$ \citep{zier05}. Such
objects might have been observed for the first time in a new sample of
about 100 XRG candidates, see \citet{cheung07}. Possibly some of these
sources exhibit broad-emission lines characteristic of quasars
\citetext{\citeauthor{cheung07}, priv.\ comm.}  as have been observed
only recently in some XRGs \citep{wang03,landt06}. This actually
strengthens the conjecture that merging \BHB{}s are the formation
mechanism for XRGs and the central torus in AGN \citep{zier01,zier02},
which is required by the unification scheme for type 1 and 2 AGN
\citep{antonucci93}: The symmetry axis of the torus, which is
surrounding the nucleus and the broad emission line region (BLR), is
aligned with the post-merger jet. Because in XRGs both lobes are close
to the plane of sky we consequently see the torus almost edge on with
the BLR hidden in its center. This is in good agreement with so few
XRGs exhibiting BLRs. The larger the angle between the plane of sky
and the post-merger jet, i.e. the axis of the torus, the more from
within the torus we can see, including the BLR, and the less reddened
the core should appear. For such objects we predict shorter
post-merger jets in projection. According to the merging scenario XRGs
are seen close to edge-on and hence they should be good candidates for
showing a type 1 spectrum in polarized light like NGC~1068
\citep{antonucci85}.

Stalled binaries are expected at the end of the second phase when
slingshot ejection of stars becomes inefficient according to some
interpretations of numerical simulations. In Section~\ref{s_ongoing},
we tested this prediction and made use of a compilation of 12 sources
\citep{merritt05} which exhibit periodic variations in their
lightcurves. These might be caused by the orbital motion of a
\BHB. From the observed periods and masses obtained with various
methods we determined the current semimajor axis of the binary,
assuming a mass ratio $q=0.1$. We find that all binaries have already
shrunk deep into the third phase and that the remaining time to
coalescence in all sources is much less than a Hubble time. The
remaining merging time increases with decreasing mass ratio and
therefore we computed $q$ for the case that it still needs a Hubble
time to merge. We obtained values much smaller than $q=0.1$ and hence
even in case of minor mergers the binary is already in the third
phase. Some of the mass estimates differed by large factors, in case
of BL~Lac by $100$. The smallest mass obtained with the method of
minimal time scales \citep{xie02} gives the only source that is still
in the second phase if $q<0.98$. However, for a larger mass which has
has been obtained with the probably more reliable $M_\mrm{BH}\,
\text{--}\, \sigma$ relation \citep{gebhardt00,tremaine02} also this
binary is clearly beyond the transition to the final phase.  Thus, in
striking contrast to the predictions of loss-cone depletion
\citep[e.g.][]{begelman80,quinlan96,makino97,quinlan97,milos01,berczik05}
we find all \BHB candidates to be already in the phase where the
emission of gravitational waves dominates the decay. Therefore a
merger with a third galaxy before the \BH{}s have coalesced and the
formation of three bound supermassive \BH{}s with the subsequent
slingshot ejection of one or more \BH{}s \citep[e.g.][]{valtonen96} is
highly unlikely.

We conclude that the \BHB which forms in a galaxy collision merges
completely.  This is in agreement with the observation of mostly
merged binaries and only few ongoing mergers with none of them being
stalled in the second phase. Hence the slingshot ejection of stars is
sufficiently effective arguing for the formation of a steep cusp at
the time when the binary becomes hard and which contains a mass of
$\sim M_{12}$, as we derived it in \citetalias{zier06} and the present
article. Triaxial potentials where the loss-cone is always full
\citep{yu02,holley-bockelmann02,holley-bockelmann06,berczik06} further
support our arguments for a successfully merged binary. The inclusion
of dark matter into our analysis, which we did not consider here
although the same formalism applies, would accelerate the merger and
make it even more likely that the \BH{}s coalesce. If there is not
enough baryonic matter in the cusp to allow the \BH{}s to merge but
they have coalesced anyway, our approach should provide a tool to draw
conclusions about the amount and distribution of dark matter in the
cusp region. However, we may conclude that stalled binaries do not
exist at all or are very rare.

\section*{Acknowledgments}
I would like to thank Wolfram Kr\"ulls for his valuable comments to
improve this manuscript. I am also grateful to the anonymous referee
for helpful advice and comments. It is a pleasure to acknowledge the
generous support and hospitality I experienced at the Raman Research
Institute. I also wish to acknowledge the hospitality of Peter
Biermann and his group at the Max-Planck Institute for Radioastronomy
while finishing the work on this paper.

\bibliography{/home/cris/LATEX/refs}
\bibliographystyle{my-mn2e}

\label{lastpage}

\end{document}